\documentclass[twocolumn]{aastex63}
\usepackage[all]{hypcap}

\graphicspath{{./}{figures/}}

\begin{document}

%\title{Heating of quiet solar corona}
%\title{Nanoflare heating of the the X-ray bright points of the quiet Sun}
\title{Role of small-scale impulsive events in heating the X-ray bright points of the quiet Sun}

\correspondingauthor{Biswajit Mondal}
\email{biswajitm@prl.res.in, biswajit70mondal94@gmail.com}

\author[0000-0002-7020-2826]{Biswajit Mondal}
\affiliation{Physical Research Laboratory, Navrangpura, Ahmedabad, Gujarat-380 009, India }
\affiliation{Indian Institute of Technology Gandhinagar, Palaj, Gandhinagar, Gujarat-382 355, India}
\author[0000-0003-2255-0305]{James A Klimchuk}
\affiliation{NASA Goddard Space Flight Center, Heliophysics Science Division, Greenbelt, MD 20771, USA}
\author[0000-0002-2050-0913]{Santosh V. Vadawale}
\affiliation{Physical Research Laboratory, Navrangpura, Ahmedabad, Gujarat-380 009, India }
\author[0000-0002-4781-5798]{Aveek Sarkar}
\affiliation{Physical Research Laboratory, Navrangpura, Ahmedabad, Gujarat-380 009, India }
%\author[0000-0002-7020-2826]{et. al}
\author[0000-0002-4125-0204]{Giulio Del Zanna}
\affiliation{DAMTP, Centre for Mathematical Sciences, University of Cambridge, Wilberforce Road, Cambridge CB3 0WA, UK}
\author[0000-0002-4454-147X]{P.S. Athiray}
\affiliation{Center for Space Plasma and Aeronomic Research, The University of Alabama in Huntsville, Huntsville, AL 35899, USA}
\affiliation{NASA Marshall Space Flight Center, ST13, Huntsville, AL, USA}
\author[0000-0003-3431-6110]{N. P. S. Mithun}
\affiliation{Physical Research Laboratory, Navrangpura, Ahmedabad, Gujarat-380 009, India }
\author[0000-0002-6418-7914]{Helen E. Mason}
\affiliation{DAMTP, Centre for Mathematical Sciences, University of Cambridge, Wilberforce Road, Cambridge CB3 0WA, UK}
\author[0000-0003-1693-453X]{A. Bhardwaj}
\affiliation{Physical Research Laboratory, Navrangpura, Ahmedabad, Gujarat-380 009, India }

\begin{abstract}

Small-scale impulsive events, known as nanoflares, are thought to be one of the prime candidates that can keep the solar corona hot at its multi-million Kelvin temperature. Individual nanoflares are difficult to detect with the current generation instruments; however, their presence can be inferred through indirect techniques such as a Differential Emission Measure (DEM) analysis. Here we employ this technique to investigate the possibility of nanoflare heating of the quiet corona during the minimum of solar cycle 24. During this minimum, active regions (ARs) were absent on the solar-disk for extended periods. In the absence of ARs, X-ray bright points (XBP) are the dominant contributor to disk-integrated X-rays. We estimate the DEM of the XBPs using observations from the Solar X-ray Monitor (XSM) onboard the Chandrayaan-2 orbiter and the Atmospheric Imaging Assembly (AIA) onboard the Solar Dynamic Observatory. XBPs consist of small-scale loops associated with bipolar magnetic fields. We simulate such XBP loops using the EBTEL hydrodynamic code. The lengths and magnetic field strengths of these loops are obtained through a potential field extrapolation of the photospheric magnetogram. Each loop is assumed to be heated by random nanoflares having an energy that depends on the loop properties. The composite nanoflare energy distribution for all the loops has a power-law slope close to -2.5. The simulation output is then used to obtain the integrated DEM. It agrees remarkably well with the observed DEM at temperatures above 1 MK, suggesting that the nanoflare distribution, as predicted by our model, can explain the XBP heating.
\end{abstract}

\section{Introduction}

Understanding the mechanism(s) that can heat the solar corona to several orders of magnitude higher than its surface ($\approx$ 6000K) remains a long-standing problem in Astrophysics. 
%heliophysics.
It is well accepted that the magnetic field lines protruding out of the photosphere
play a crucial role in heating the corona.
The footpoints of the field lines are randomly moved by the convective motions below the photosphere, causing either the quasi-static build up of magnetic stress or the generation of waves depending on the time scales of the motion~\citep{Klimchuck_2006SoPh}. %which is finally get converted to heat.
Dissipation of magnetic stress is known as DC heating whereas the dissipation of waves is known as AC heating.
Most of the models of coronal heating, both DC and AC, suggest that the heating is impulsive in nature~\citep{Klimchuck_2006SoPh}.
\cite{klimchuk_2015RSPTA} defines the small-scale impulsive events as nanoflares irrespective of the underlying physical mechanism.
The magnitude and occurrence frequency of these nanoflares determine whether they can provide sufficient energy required for the total heating.
Thus, it is of great importance to investigate the likely magnitudes and frequencies of nanoflares to validate the impulsive heating models.

Due to line-of-sight averaging and the finite spatial resolution of the present generation of instruments, direct observation of the nanoflares is difficult. 
Instead of their direct observable signature several indirect methods are used to infer their existence, e.g., `Intensity Fluctuations'~\citep{Katsukawa_2001ApJ,Pauluhn_2007,Sakamoto_2008ApJ...689.1421S}, `Time Lags'~\citep{Viall_2012ApJ...753...35V,Viall_2013ApJ...771..115V,Viall_2015ApJ...799...58V,Bradshaw_2016ApJ...821...63B}, `differential emission measure' (DEM) or the `emission measure distribution' (EMD). The DEM gives an estimation of the amount of plasma present at different temperatures (per unit temperature) and the integration of DEM over temperature bins provides  the EMD.
%This provides an indirect signature of the energy deposition via observations of the plasma cooling by thermal conduction, enthalpy, and radiation~\citep{Barnes_2019,Hinode_review_2019PASJ}.

The DEM technique has been extensively used in many observational studies to interpret the heating of quiescent  active region core in terms of heating frequencies
(e.g., ~\citealp{Tripathi_2011,Winebarger_2011,giulio_2015A&A,Brosius_2014ApJ, Caspi_2015,Ishikawa_2017NatAs}).
However, use of this technique is rare to study the quiet Sun heating.
Earlier studies of the quiet Sun DEM
~\citep{Lanzafame_2005A&A,Brooks_2009,Giulio_2019A&A} show a peak at low temperatures, around 1 MK. 
However, determining the DEM for the quiet Sun at high temperatures ($> 2$ MK) turns out to be difficult because of the faint emission at this temperature range~\citep{Giulio_2018LRSP}.
Lately, using Hard X-ray observations, \cite{Paterson_2022} derived DEMs for different features of the quiet solar corona. They found faint emission up to $4$ MK.

In the present study, we derive the quiet Sun DEM
 using Sun as a star observations during the minimum of solar cycle 24. 
 Here, the quiet Sun includes the quiet diffuse regions (defined as QDR), the so-called diffuse corona emitting in the temperature $\sim$ 1 MK; cool coronal holes which mostly emit at a lower temperature ($<$ 1 MK); the X-ray emitting regions (XER), which are the origin of most of the X-ray emission including the limb brightening and X-ray bright points (XBP).
 We use combined observations in soft X-rays and Extreme Ultraviolet (EUV) energy bands from the Solar X-ray Monitor (XSM:~\citealp{vadawale_2014,shanmugam_2020}) onboard Chandrayaan-2 orbiter and Atmospheric Imaging Assembly (AIA:~\citealp{Lemen_2012SoPh}) onboard Solar Dynamic Observatory.
The XSM observations during the minimum of solar cycle 24 were used earlier to study the quiet solar corona using an isothermal assumption~\citep{xsm_XBP_abundance_2021}. 
 Comparing the X-ray images of the Sun by the Be-thin filter of XRT/Hinode, 
 %XRT Be-thin filter [**JK: define. Hinode not yet mentioned.], 
 whose high energy response is similar to the XSM lower energy response, \cite{xsm_XBP_abundance_2021} inferred that a large fraction ($>$ 50\%) of the quiet Sun X-ray emission arises from the X-ray Bright Points (XBP). They derived the isothermal temperature, emission measure, and elemental abundances for XBPs. 
 In the present study, we have extracted the contribution of XER and then XBPs from the total quiet Sun emission to estimate their DEM separately. 
 We quantify the emission from the XER and XBPs to the total X-ray emission.

XBPs consist of small-scale rapidly evolving coronal loops~\citep{Madjarska_2019}. 
Using the Enthalpy-Based Thermal Evolution of Loops (EBTEL:~\citealp{klimchuck_2008ApJ, cargill_2012a, cargill_2012b}) model, we simulate the XBP loops and determine their DEM. We derive a composite DEM for all the loops and compare this with the observations. 

The frequency distribution of the impulsive events, so-called nanoflares, for which the simulated DEM of XBPs matches the observation is further compared with the frequency distribution of the microflares as observed by  XSM ~\citep{xsm_microflares_2021} in the quiet Sun.  These microflares have energies $\sim3\times10^{26}-6\times10^{27}$ erg, and most of them were found to be associated with XBPs. 

The rest of the paper is organized as follows.  In Section~\ref{sec-observation} the observation and the data analysis of the XSM and AIA are presented. In Section~\ref{sec-Combined_EM_analysis} the detailed method of the combined DEM analysis and results are described.
Description of the XBPs simulation setup and results are given in Section~\ref{sec-simulations}. 
Finally, we discuss and summarize the primary findings of the work in Section~\ref{sec-discussion}.

\section{Observations and Data Analysis}\label{sec-observation}

\begin{figure*}[ht!]
\centering
\includegraphics[width=1\linewidth]{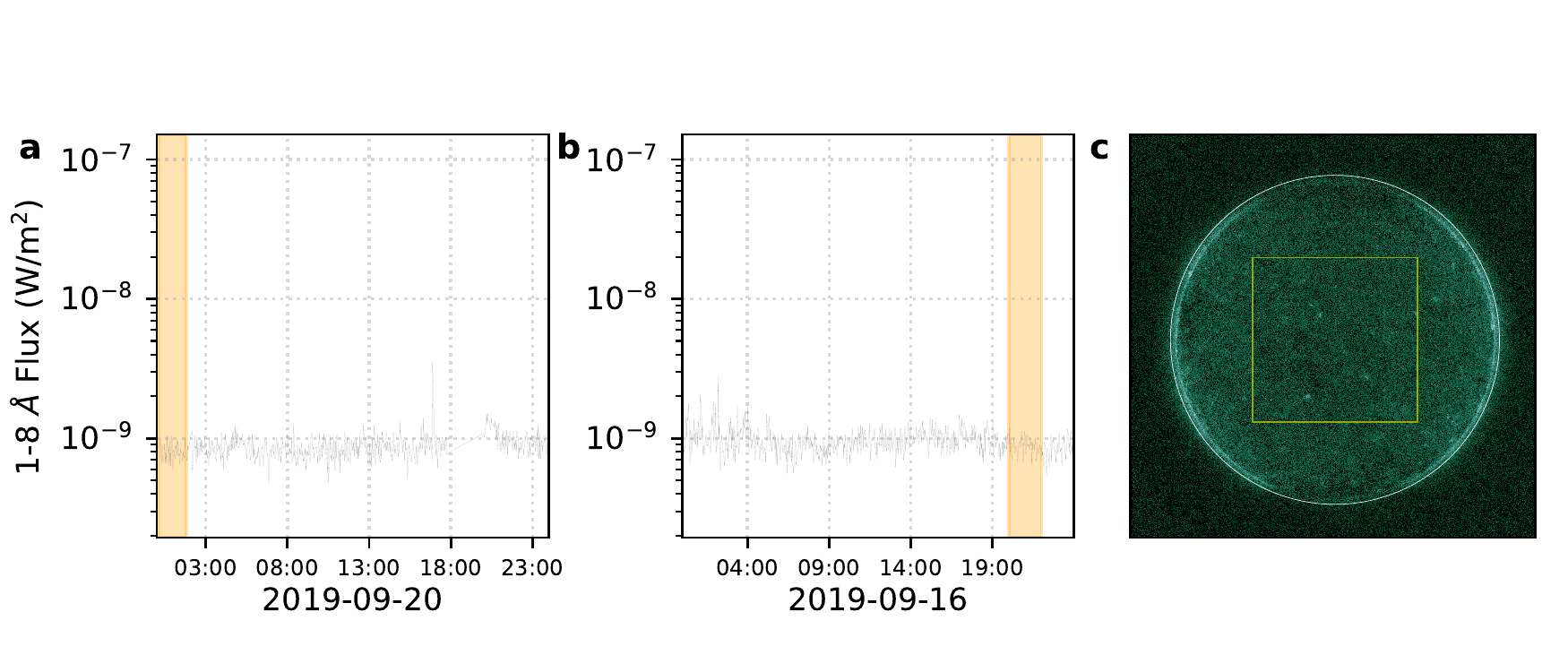}
\caption{Panels a and b show the 1-8\textup{\AA} light curve of the Sun observed by XSM during Sep 20 and Sep 16 2019. The orange shaded region represents the duration of QS-1 and QS-2 as mentioned in the text. Panel c shows a representative full disk image of the Sun during QS-1 taken by AIA 94\textup{\AA} channel. The yellow square box at the centre shows 1000$\arcsec \times$ 1000$\arcsec$ field-of-view as mentioned in Section~\ref{sec-XBP_EMdist}.}
\label{fig-LC}
\end{figure*}

We use the X-ray observation of the Sun by XSM
onboard India's Chandrayaan-2 orbiter.
XSM measures the disk integrated solar spectra in the energy range of 1-15 keV at every second with an energy resolution better than 180 eV at 5.9 keV~\citep{shanmugam_2020,XSM_ground_calibration}.
The unique design of XSM makes it possible to observe a wide range of solar X-ray intensities from the quiet Sun to X-class solar flares~\citep{xsm_flight_performance}. 
XSM started solar observations on September 12, 2019, and observed well the minimum of solar cycle 24 covering the years 2019 and 2020. 
During September 2019 to May 2020, there were 76 days when no active regions (AR) were present on the solar disk ~\citep{xsm_XBP_abundance_2021}, defined as the quiet-Sun (QS) period. 
In the present study two representative intervals are selected from the QS duration on September 20, 2019 (00:07 UTC - 01:49 UTC, defined as QS-1) and September 16, 2019, (20:00 UTC - 22:00 UTC, defined as QS-2).
Following the standard analysis procedures described in~\cite{xsm_XBP_abundance_2021} or~\cite{biswajit_2021}, we generate the XSM observed flux light curves in the energy range of 1-8$\textup\AA$ for the days that include QS-1 and QS-2 as shown in  Figure~\ref{fig-LC}a,b. The orange shaded color marks the duration of QS-1 and QS-2.

The primary objective of the present study is to estimate the DEM/EMD for the QS period.
Since XSM is more sensitive to higher temperatures ($>$2 MK), to constrain the DEM at lower temperatures ($<$ 2 MK) we need to combine XSM with EUV data (see Section~\ref{sec-Combined_EM_analysis}).
%In the present study, we are interested in estimating the EM as a function of temperature for the coronal plasma during the QS period, which requires multi-wavelength observation from EUV to X-rays. 
Thus, we combine the EUV observation from AIA on
board Solar Dynamics Observatory. 
AIA continuously records full-disk images of the Sun in different EUV energy channels (94 $\textup\AA$, 131 $\textup\AA$, 171 $\textup\AA$, 193 $\textup\AA$, 211 $\textup\AA$, 304 $\textup\AA$, 355 $\textup\AA$) with a cadence of 12s.
During the QS-1 and QS-2 periods, the level-1 AIA full-disk images in all of its pass-bands were downloaded from  Joint Science Operations Center (JSOC) and processed to level-1.5 using the standard routines available in the SolarSoftWare package (SSW; \cite{Freeland_1998}). 
A representative full-disk image frame of the AIA 94 $\textup\AA$ channel during the QS-1 period is shown in Figure~\ref{fig-LC}c.

\section{Combined DEM Analysis}\label{sec-Combined_EM_analysis}

The  DEM or EMD gives an indication of the amount of plasma that is emitting the radiation observed, and has a temperature between $T$ and $T+dT$~\citep{Giulio_2018LRSP}.
To estimate the DEM we use simultaneous observatios at several  EUV and X-ray energy bands,  sensitive to different temperatures.
We use the five EUV channels of AIA, 94 $\textup\AA$, 131 $\textup\AA$, 171 $\textup\AA$, 193 $\textup\AA$, and 211 $\textup\AA$ that are sensitive to temperatures more than $log T = 5.6$. 
We exclude the channel 335 $\textup\AA$ due to a long-term drop in sensitivity resulting from accumulated contamination~\citep{Boerner_2014SoPh,Athiray_2020}. 
For each AIA channel, we consider the integrated intensity of all the positive finite pixels below a solar radius of 1.04 R$_\odot$ (white circle in Figure~\ref{fig-LC}c), from where most of the emission is coming in all the energies. 
We have verified that the final results remain unaffected even if we consider the pixels within a larger radius or even the full AIA Field-Of-View (FOV). 
For the X-ray observation,
we divide the XSM spectrum into four energy channels  of $1.29-1.45$ keV, $1.45-1.75$ keV, $1.72-1.95$ keV, and $1.95-2.5$ keV. 
These channels were chosen such that each includes a line complex of particular element/elements (Mg, Mg+Al, Si, and Si$+$S) with good statistics.
Thus we obtain the observed intensity in a total of nine instrument channels, five channels from AIA, and four channels from XSM.

The Observed intensity ($O_i$) at i'th instrument channel is related to the DEM as follows:
\begin{equation}\label{eq_dem}
O_i = \int_{T} DEM(T)\hspace{0.1cm} R_{i}(T)\hspace{0.1cm} dT \hspace{0.1cm} + \hspace{0.1cm}\delta O_i
\end{equation}  
Here, $\delta O_i$ is the uncertainty associated with $O_i$, and $R_i(T)$ is the temperature response function of the i'th channel. A  temperature response represents the sensitivity of an instrument channel to detect the plasma emission at different temperatures. 
Figure~\ref{fig-AIA_XSM_Tresp} shows the temperature response functions for AIA channels (dashed lines) along with the four XSM channels (solid lines).
The detailed method to obtain the temperature response functions of XSM and AIA is described in Appendix~\ref{XSM_Tresp}.

\subsection{Full Sun DEM ($DEM_{\mathrm{FullSun}}$)}\label{sec-fullSunEM}

The observed intensities ($O_{i}$) and temperature response functions of all the energy channels are already known. To recover the DEM we use the \verb|xrt_dem_iterative2.pro|~\citep{Golub_2004ASPC} method (say \verb|xrt_dem}|). This is basically a forward-fitting routine which finds the DEM solution from Eq.~\ref{eq_dem} by considering a spline function for the DEM curve.
This routine is a standard tool-set for solar data analysis in the SolarSoftWare (SSW; \cite{Freeland_1998}) package.
The best-fit DEM is identified iteratively using a nonlinear least-square method by comparing the predicted and observed fluxes. 
This method has been widely used in DEM fitting with AIA/SDO, XRT/Hinode, EIS/Hinode, 
and FOXI data (e.g., \cite{Golub_2007SoPh,Winebarger_2011,Ishikawa_2017NatAs, Wright_2017,Athiray_2020}).
Here, we consider a temperature range of 5.9 $\le$ log$T$ $\le$ 6.8 with a bin size of $\delta$log$T$ = 0.03 for the DEM estimation.
The uncertainties in the recovered DEM are estimated through Monte-Carlo (MC) runs, which are performed by varying the observed intensities randomly within the observed errors. 
The errors in the AIA observed counts at each channel are estimated using the standard procedure, \verb|aia_bp_estimate.pro|~\citep{Boerner_2012SoPh}. 
Uncertainties in the XSM observation primarily contain the Poisons error associated with the counting statistics and small systematic errors at each spectral channel provided by the XSM data processing software.
To estimate the uncertainties in the recovered DEM solution,
we perform a large set (500000) of MC runs over the observed counts.
Among  all the MC samples, we ignore the spurious DEM solutions, e.g., selecting the DEM solutions that can describe the observed flux at all channels with a reduced-chi square of less than equal to 2. 
The histogram of the DEMs at each temperature node is derived using the accepted DEM solutions.
From the peak of the DEM histogram at each temperature node, we estimate the one-sigma uncertainties.

\begin{figure*}[ht!]
\centering
\includegraphics[width=1\linewidth]{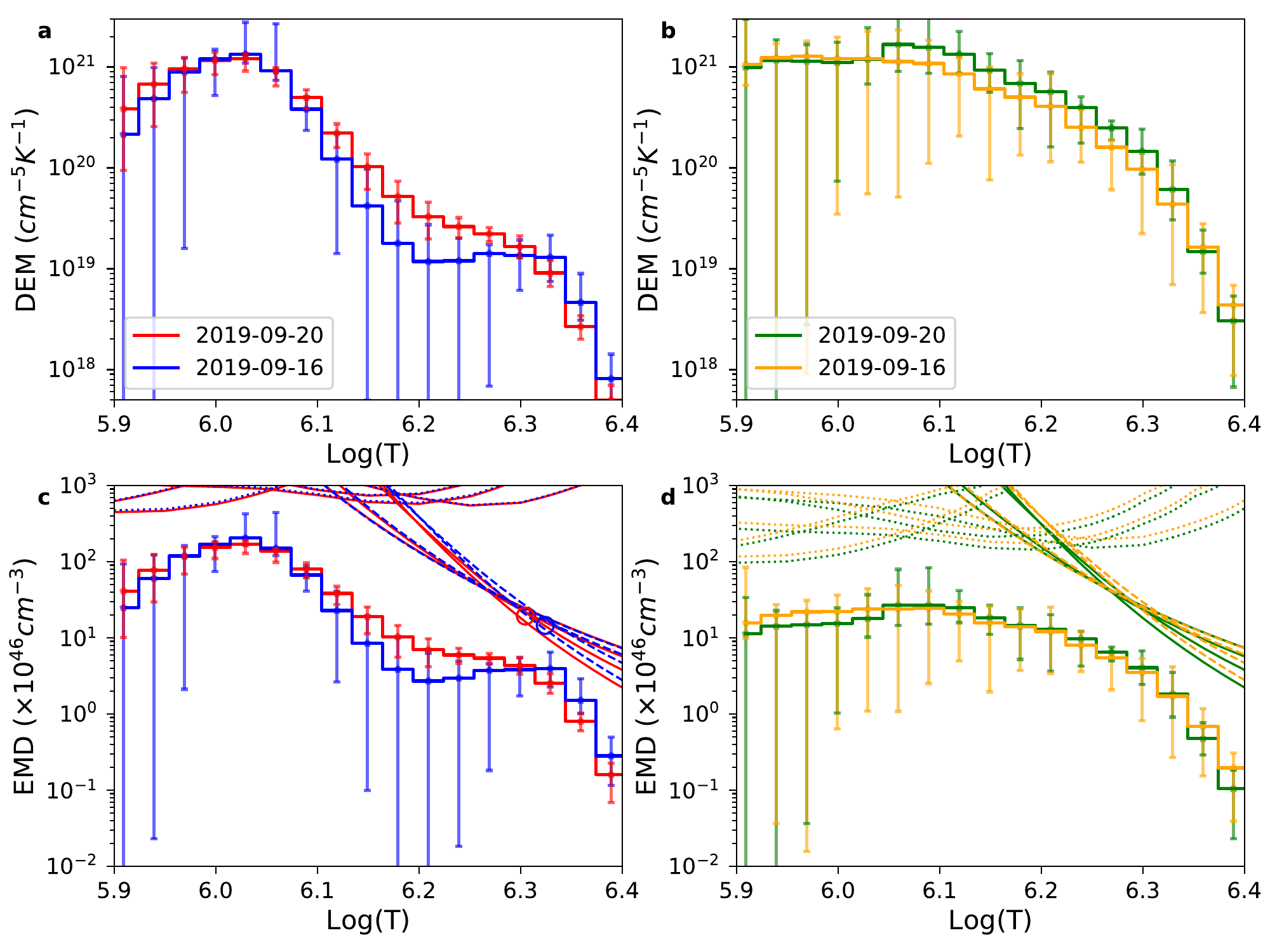}
\caption{Panels {\bf a} and {\bf c} show the full Sun DEM and EMD profile for QS-1 (red) and QS-2 (blue). Panels b and d show the DEM and EMD for the XERs associated with QS-1 (green) and QS-2 (orange). 
The EM loci curves for AIA (dashed lines) and XSM (solid lines) are overplotted in Panels {\bf c} and {\bf d}. The red and blue circular points in Panel {\bf c} represent the isothermal temperature and EM for the XER as reported by~\cite{xsm_XBP_abundance_2021}}
\label{fig-EM_distribution}
\end{figure*}

The full-Sun DEM (defined as $DEM_\mathrm{FullSun}$) and the 1-sigma error bars are shown in  Figure~\ref{fig-EM_distribution}a for QS-1 (red) and QS-2 (blue). 
The solid line represents the peak of the DEM histogram at each temperature node.
We derive the EMD (units of cm$^{-3}$) from DEM (units of cm$^{-5}$k$^{-1}$) by multiplying the DEM with $area \times T\delta logT$ (here, $area$ is the total emitting area on the Sun and $\delta logT$ is the logarithmic bin size of temperatures). 
Derived EMD for QS-1 and QS-2 are shown in Figure~\ref{fig-EM_distribution}c.
Dividing the observed counts with the temperature response function associated with each channel gives the emission-measure loci curves, which indicate the upper limit of the EMD.
The emission-measure loci curves for the QS-1 (red curves) and QS-2 (blue curves) at the five AIA channels (left side) and four XSM channels (right side) are overplotted. 

Note that here we assume an integrated emission from the AIA images which includes the emission from the quiet region, XBPs, and the limb emission. 
However, from the full disk X-ray images (e.g., XRT/Hinode Be-thin filter images) one can see that the X-ray emission from the quiet region is very very faint compared with the XBPs  and limb emission.
Thus, in the next step we separate the X-ray Emitting Regions (XER) from the AIA full-disk images as discussed in Section~\ref{appendix-XbpDetec} and then combine the intensity of XER from AIA images with the XSM observation to estimate the combined DEM of the XER, as discussed in Section~\ref{sec-XBP_EMdist}.

\subsection{Identification of XER in AIA EUV images}\label{appendix-XbpDetec}

In the full-disk X-ray and EUV images of the Sun, the XER are 
 found to be bright compared with the surrounding quiet Sun emission. 
Thus, the XER emission can be separated out using a source detection technique.
In this study, we have used the astronomical source detection algorithm, \verb|Photutils|~\citep{Photutils}
over the full-disk image of the AIA 193 $\textup\AA$ channel to estimate the typical 
emitting regions of the XER.
$Photutils$ is a Python library that provides tools for detecting astronomical sources using image segmentation. 
The detected sources must have a minimum number of adjacent pixels, each of them greater than a given threshold value in an image.
Usually, the threshold value is taken to be the background noise (sigma) multiplied by a factor.
In our case, we have estimated the background noise of the quiet Sun emission in the AIA 193 $\textup\AA$ images using the \verb|detect_threshold| method of the \verb|Photutils| and defined a threshold level of two times the background noise.
We apply a 2D circular Gaussian kernel with a Full-Width-Half-Maximum (FWHM) of three pixels to smooth the image prior to applying the threshold.
Using the \verb|detect_sources| method of the Photutils we find out all the distinct sources that have a minimum of five connected pixels. 
A mask frame of the same dimension as the original image is prepared by assigning all the detected source pixels a value of one and the rest a value of zero. 
Multiplying the mask with the original image gives us a mask image, 
which provides the typical contribution of the XER. The same mask frame is used in all the AIA channels to find out the XER contribution in the respective pass-band.

\begin{figure}[ht!]
\centering
\includegraphics[width=1\linewidth]{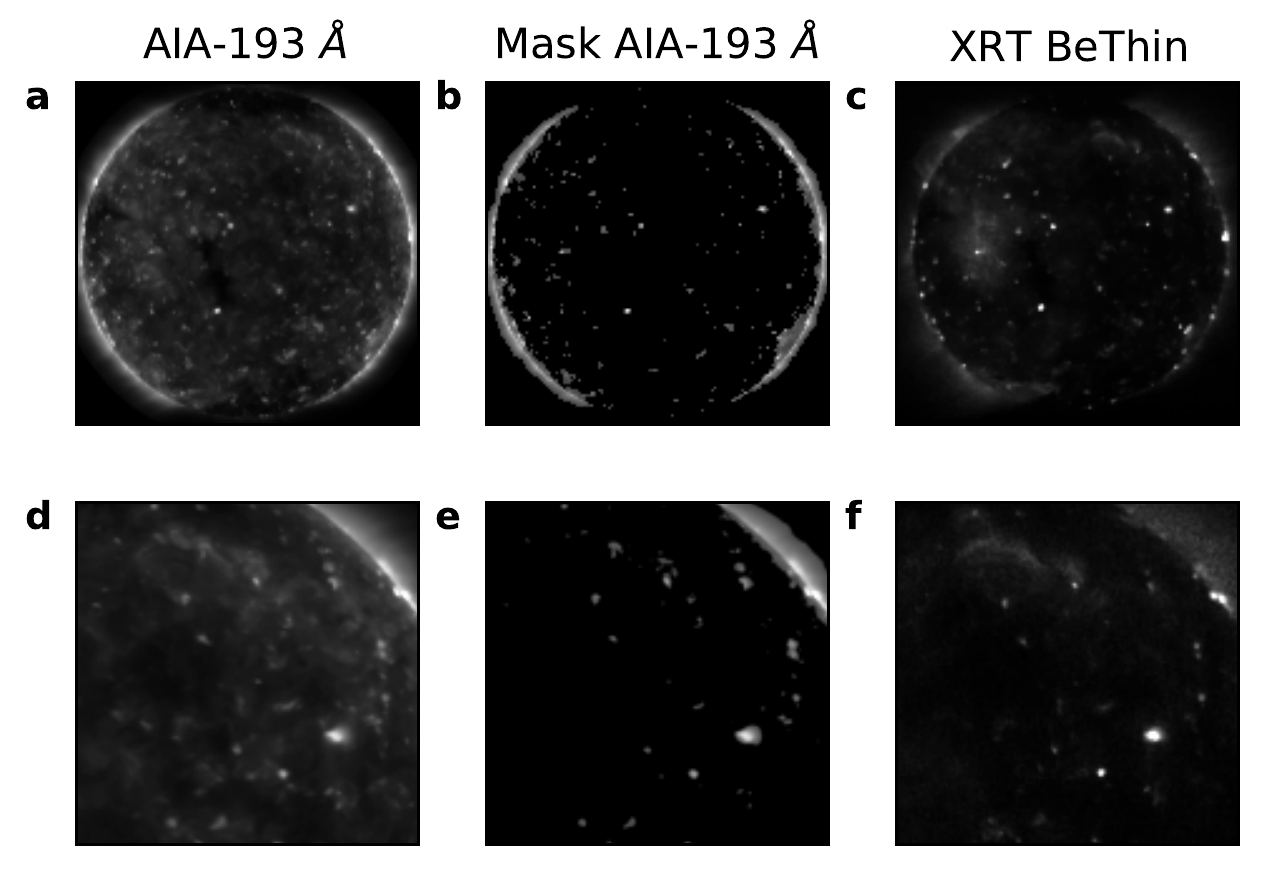}
\caption{Full disk images of the Sun during QS-1 taken by AIA 193\textup{\AA} channel (Panel {\bf a}) and XRT Be thin filter (Panel {\bf c}). Panel {\bf b} shows the XERs extracted from the AIA 193\textup{\AA} image. Panels shown in the bottom row represent a portion of the solar disk taken from the above panels.
}
\label{fig-XBP_detection}
\end{figure}

Panels a and d in Figure~\ref{fig-XBP_detection} show a representative full-disk solar image and a zoomed view of the same image  on 20-09-2019, taken in the AIA 193 $\textup\AA$ channel.
The bright regions represent the emission from XER.
Panels b and e show the masked images of the original images (panels a and d).
The masked images show a good agreement with the X-ray images taken by the XRT Be-Thin filter as shown in panels c and f.
The emission in the masked images (panels b and e) is well matched with the X-ray images (panels c ad f) except for some negligible portions here and there. The limb emission is not as noticeable in X-rays as it is at 193 $\textup\AA$, as we discuss later.

\subsection{DEM of XER ($DEM_\mathrm{XER}$)}\label{sec-XBP_EMdist}

Using the integrated emission from the XER in the AIA images (Section~\ref{appendix-XbpDetec}) along with the X-ray emission detected by XSM, we derive the DEM of X-ray emitting regions (define as $DEM_\mathrm{XER}$) in a  similar manner to the full-Sun DEM (Section~\ref{sec-fullSunEM}).
Panels b and d of Figure~\ref{fig-EM_distribution} show the DEM and EMD of the XER during QS-1 (green) and QS-2 (orange). EM-loci curves for all AIA channels (left) and XSM channels (right) are also shown in panel d.
In the full Sun, as the emitting area for the high temperature ($>$1.5 MK) emission is less than the cool plasma, the full Sun DEM (panel a) at high-temperatures is much lower than that of the XER (panel b).
However, comparing the EMD of the full Sun (panel c) and the XER (panel d), we can say that the higher temperature ($>$ 1.5 MK) portion is similar. This indicates that the hotter emission comes primarily from the XER. At lower temperatures, where the EMD is primarily determined by AIA, the full Sun EMD shows an excess emission. 

\subsection{Validation of recovered DEMs}

\begin{figure*}[ht!]
\centering
\includegraphics[width=1\linewidth]{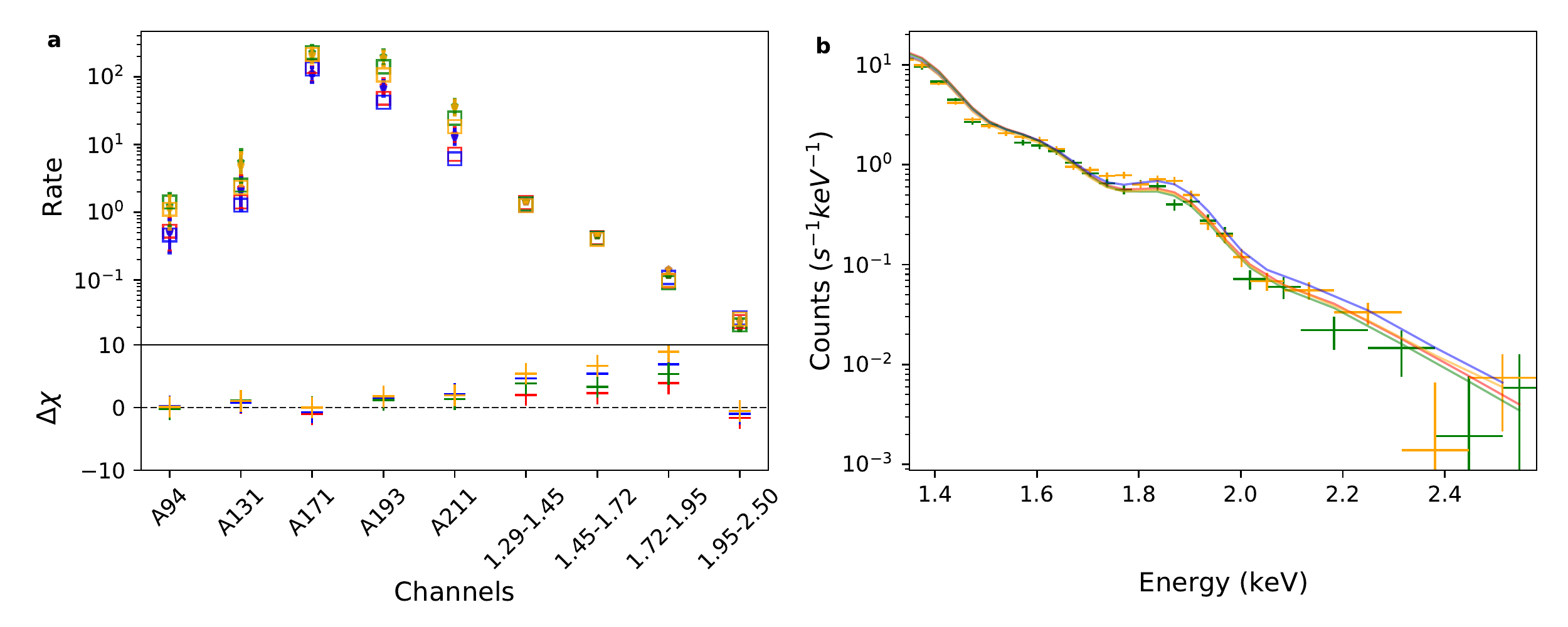}
\caption{Panel {\bf a} represent the observed intensities (in \textit{DN Px$^{-1}$ s$^{-1}$} for AIA and \textit{Counts s$^{-1}$} for XSM) of QS-1 and QS-2 measured in the different channels of AIA and XSM and the square boxes represent the predicted intensities using the DEM shown in Figure~\ref{fig-EM_distribution}. The panel underneath shows the delta-chi (i.e., (observed-predicted)/error) between the observed and predicted intensities. The error bars in Panel {\bf b} show the observed XSM spectra of QS-1 and QS-2. The solid lines shown by red, blue, green, and orange colors represent the predicted XSM spectra using the DEM shown in Figure~\ref{fig-EM_distribution}.}
\label{fig-Pred_Obs_intensity}
\end{figure*}

To verify the reliability of the recovered DEM/EMD as discussed in Sections~\ref{sec-fullSunEM} and \ref{sec-XBP_EMdist}, we estimate the predicted counts in all channels and the XSM spectra by using the recovered DEM and then compared them with the observed intensities and XSM spectra.
The top panel of Figure~\ref{fig-Pred_Obs_intensity}a shows the observed (points with error bars) and predicted (box points) intensities (same color as Figures~\ref{fig-EM_distribution}) in all the instrument channels using the DEM shown in Figure~\ref{fig-EM_distribution}. 
The bottom panel indicates the delta-chi ((Observed-Predicted) / Error) between the observed and predicted intensities, where the errors are the uncertainties in the observed intensities, $\delta O_i$ in Equation 1. 
The predicted intensities for all the recovered DEMs match the observed intensities to within the error bars.

Further, we forward-model the XSM spectra using the recovered EMD of both the full Sun (blue and red solid lines correspond to QS-1 and QS-2) and XER (orange and green solid lines correspond to QS-1 and QS-2), as shown by solid lines in Figure~\ref{fig-Pred_Obs_intensity}b. For comparison, the observed XSM spectra during QS-1 (green error bars) and QS-2 (orange error bars) are overplotted.
The modeled spectra derived from the EMD agree well with the observed ones.
As XSM is most sensitive to higher temperatures, the excess emission at lower temperatures in the full-Sun EMD (Figure~\ref{fig-EM_distribution}c) does not contribute much to the modeled XSM spectra. Thus, the EMD from the full Sun and X-ray emitting regions explain the XSM spectra equally well. This verifies that most of the emission observed by XSM primarily originates from X-ray emitting regions.

\subsection{DEM of XBPs ($DEM_\mathrm{XBP}$)}\label{sec:XBP_dem}
The DEM of the XER has contributions from both the XBPs and the limb brightening. Though the limb seems to be very bright in the full-disk images (Figure~\ref{fig-XBP_detection}; specifically in AIA energy bands), it is well known that the limb emission primarily comes from cool plasma of great line-of-sight depth. Thus, it is expected that the limb emission contributes mostly to the lower-temperature part of the DEM, whereas the high-temperature part comes primarily from the XBPs.
However, in our recovered DEM we found that at lower temperatures the error bars are very large and we could not predict DEM at very low temperatures, e.g., logT $<$ 5.9. This is due to the fact that at those temperatures the emissions are very faint and hence noisy, reliable results could not be recovered by the \verb|xrt_dem| method.
This uncertainty has been demonstrated nicely by~\cite{Hannah_2012} for a set of simulated data of AR and quiet Sun for different AIA channels. 
Using a regularized inversion to solve Equation~\ref{eq_dem}, 
\cite{Hannah_2012} gave a different approach to estimate the DEM from the observed intensity of different instrument channels.
The major advantage of this method is that it determines the errors of estimated DEM along with the uncertainty in temperature intervals.
In the next step, we apply the  \cite{Hannah_2012} method\footnote{https://github.com/ianan/demreg/tree/master/python} (define as \verb|HK_dem|) to recover the DEM and compare the obtained DEM with that obtained by the \verb|xrt_dem| method. 

\begin{figure*}[ht!]
\centering
\includegraphics[width=1\linewidth]{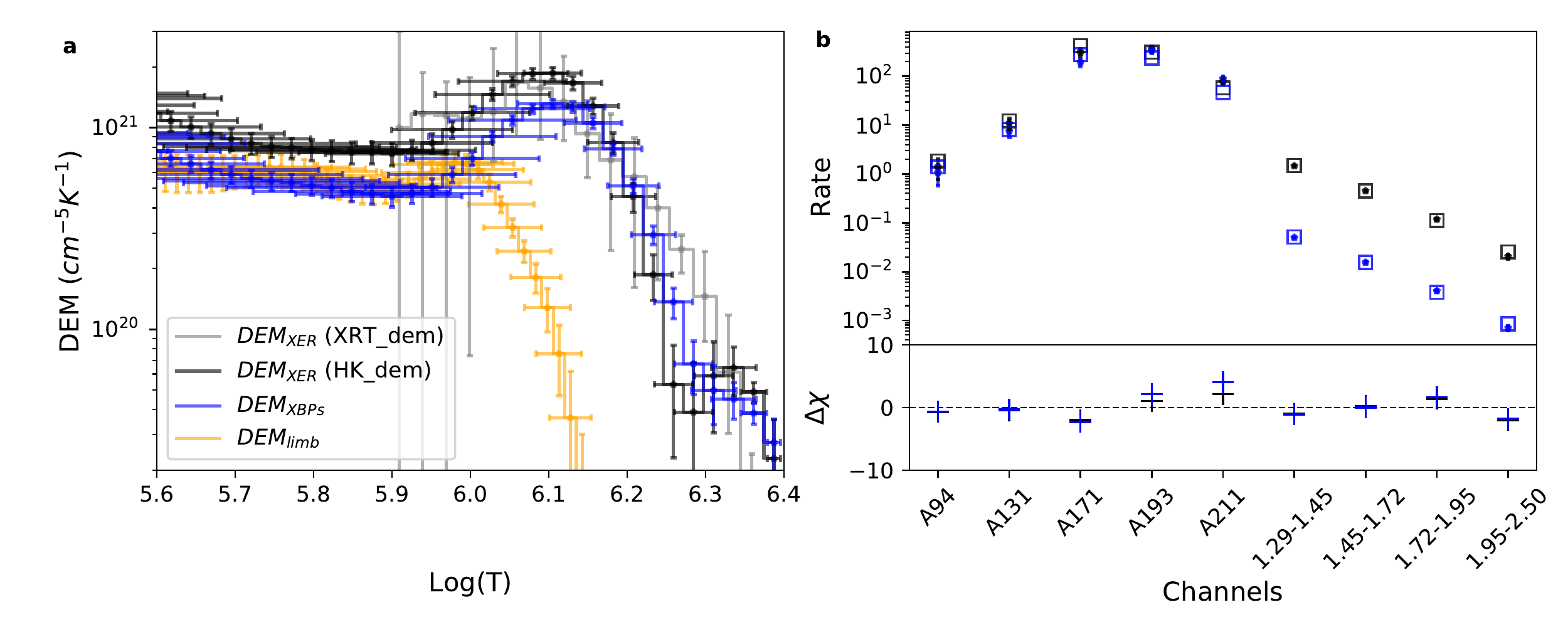}
\caption{Panel {\bf a} shows the observed DEM of XER (black), XBPs (blue), and limb (orange) derived by $HK\_dem$ method. The DEM of XER derived by $XRT_\mathrm{dem}$ method is overplotted by grey color. Panel {\bf b} shows the observed (error bars) and predicted counts in different instrument channels.}
\label{fig-HK_dem}
\end{figure*}

Using the \verb|HK_dem| method we have recovered the $DEM_\mathrm{XER}$ of QS-1 down to a lower temperature (log(T) = 5.6).
Figure~\ref{fig-HK_dem}a shows the derived DEM in both the \verb|HK_dem| and \verb|xrt_dem| method.
Both the methods provide very similar results at higher temperatures ($>$ 1MK). The \verb|HK_dem| provides the  DEM at lower temperatures with very large error in the log(T) resolution, which could underestimate the lower temperature DEM in a similar way to that demonstrated by~\cite{Hannah_2012}. 

Our objective here is to extract the DEM of the XBPs (define as $DEM_\mathrm{XBP}$) located within an area of
1000$\arcsec$ $\times$ 1000$\arcsec$ (say $F_v$) at disk center (yellow box in Figure~\ref{fig-LC}c). 
This would be very straightforward if we knew the counts detected by the XSM only from the XBPs located inside the $F_v$.
But, the emission detected by XSM includes XBPs from the whole disk as well the limb emission. Quiet Sun emission is negligible, as we have discussed. We can estimate the XSM emission from XBPs inside $F_v$ by degrading the total XSM counts by a factor $f$. In a first attempt, we estimate $f$ as the ratio of the number of XBPs inside $F_v$ and that of the whole disk. However, this overestimates the emission from the XBPs inside $F_v$ because it ignores the limb emission. The $DEM_\mathrm{XBPs}$ solution is therefore spurious.
A better approach is to estimate $f$ as the ratio of the area of XBPs inside $F_v$ and the area of total XER, which we obtain from our masked image, Figure~\ref{fig-XBP_detection}c. This provides a reasonable solution of the $DEM_\mathrm{XBP}$.
Blue error bars in Figure~\ref{fig-HK_dem}a show the estimated $DEM_\mathrm{XBP}$ and this DEM predicted the observed intensities very well (Figure~\ref{fig-HK_dem}b). 
The $DEM_\mathrm{XBP}$ differs from the DEM of total XER only at lower temperatures ($< $1 MK), which is expected as at lower temperatures the limb emission contributes to the total DEM. 

A more sophisticated verification of contribution of limb emission to the total is done by estimating the typical DEM of the limb (say $DEM_\mathrm{limb}$) emission using the different channels of AIA along with the XRT filter images. 
We select a small portion of the limb and then estimated the counts in AIA EUV channels along with the XRT \verb|Al-mesh|, \verb|Al-poly|, \verb|Be-thin| filters.
Using a 20$\%$ uncertainty with the observed intensity along with a calibration factor of 2~\citep{Athiray_2020ApJ} for XRT, we estimated the $DEM_\mathrm{limb}$.
The recovered DEM for the limb is shown in orange color in Figure~\ref{fig-HK_dem}. 
This also indicates that the limb is only contributing emission at a lower temperature.

It should be noted that at temperatures below 1 MK, there may be some uncertainty in determination of f, and hence in $DEM_\mathrm{XBP}$; however, at temperature above 1 MK the estimated $DEM_\mathrm{XBP}$ is quite robust, as the contribution of the limb emission at these temperature is negligible. Thus it can be safely assumed that the $DEM_\mathrm{XBP}$ shown in Figure~\ref{fig-HK_dem} represents the \textit{average} DEM for the XBPs. 

\section{Hydrodynamic Simulations of XBPs emission}\label{sec-simulations}

To investigate the energy requirement for XBPs to maintain the observed DEM ($DEM_\mathrm{XBP}$), we carry out hydrodynamic simulations.
XBPs are found to be associated with bipolar magnetic field regions, similar to active regions,  and consist of independent, rapidly evolving small-scale loops~\citep{Madjarska_2019}. It is thus natural to assume that the hot emission of XBPs is associated with the confined plasma within the small-scale magnetic loop systems, termed a magnetic skeleton.
Field-aligned hydrodynamic models are often used to estimate the evolution of the plasma confined within the coronal loops. One such model is  Enthalpy-Based Thermal Evolution of Loops model (EBTEL;~\cite{klimchuck_2008ApJ, cargill_2012a, cargill_2012b}). EBTEL is a zero-dimensional (0D) time-dependent hydrodynamic model that can accurately estimate the time evolution of the spatially averaged coronal temperature, density, and pressure of a single coronal loop heated by an assumed heating profile (time-dependent heating rate).
The primary advantage of using 0D models such as EBTEL for such simulations is that their run time is orders of magnitude faster than that of spatially resolved 1D models.  
%The primary advantage of using EBTEL is its simulation timescale, which is an order of magnitude faster than the spatially resolved 1D models.
Despite the simplicity of the EBTEL calculation, it can provide plasma parameters very similar to the loop-averaged values from 1D models.
Along with the average coronal properties of the loop, EBTEL estimates the DEMs of the transition region and coronal portion of the loop separately at each time step. 
In this work we have used the two-fluid version of EBTEL (EBTEL\footnote{https://rice-solar-physics.github.io/ebtelPlusPlus/}++:~\cite{Barnes_2016a, Barnes_2016b}), where the ions and electrons are treated separately; a detailed implementation of it can be found in~\cite{Barnes_2016a}.

Using the high resolution full-disk photospheric magnetic field measurements from the Helioseismic and Magnetic Imager (HMI:~\cite{scherre_2012SoPh}) onboard the SDO, we have extrapolated the magnetic field lines and produced the magnetic \textit{skeletons} associated with the XBPs as discussed in Section~\ref{sec-extrapolation}, which provide the lengths and magnetic field strengths of the loops that comprise the skeleton. The loops are simulated with EBTEL using heating profiles that depend on the length and field strength as described in Section~\ref{sec:heating_function}. The approach is similar to that used by~\cite{Nita_2018} for active regions. 

\subsection{Magnetic skeleton of XBPs}\label{sec-extrapolation}

We are interested in modeling all the XBPs emissions near the disk center within an area of  1000$\arcsec$ $\times$ 1000$\arcsec$ (defined as $F_v$ in Section~\ref{sec:XBP_dem}).
Using the locations of all the XBPs within $F_v$ (Section~\ref{appendix-XbpDetec})
we identified their counterpart on the full-disk line-of-sight (LOS) HMI magnetogram and find that all of them are associated with magnetic bipolar regions.
Considering these bipolar regions as a lower boundary, 
we extrapolate their field lines up to a height of 200 HMI pixels (~72 Mm).
For this purpose, we use the Linear Force-Free Extrapolation code, \verb|j_b_lff.pro|~\citep{nakagawa_1972,Seehafer_1978SoPh}, available within the SSW by setting the force-free parameter $\alpha$ $=$ 0. The field is therefore a potential field. 
Using the three-dimensional extrapolated magnetic fields data, we trace field lines through the volume corresponding to the XBP following the streamline tracing method. 
For the streamline tracing we have chosen the seed points (to which, field lines are passed through) randomly within the extrapolated volume.

We assume that each traced field line corresponds to a coronal loop, and the loop has a constant radius ($r$) of 1 Mm throughout the height. 
The number of loops associated with an XBP is found by divided the total area of the XBP in the masked AIA image by the combined cross sectional area of the two footpoints. If the area of $i^{th}$ XBP is $A_i$, then it contains $N_i$ loops, where
\begin{equation}\label{eq_loopNo}
    N_i = \frac{A_i}{2\pi r^2}
\end{equation}

Using the coordinates ($x_k$, $y_k$, $z_k$) and magnetic field strength ($B_{x_k}$, $B_{y_k}$, $B_{z_k}$) of each of the loop along their length, we derive their length ($L$) and  average magnetic field strength ($<B>$) as follows:
\begin{equation}
   L = \sum_{k}\sqrt{((x_{k+1}-x_k)^2 +  (y_{k+1}-y_i)^2 +(z_{k+1}-z_k)^2)} ~
\end{equation}

\begin{equation}
   <B> = \frac{\sum_k\sqrt{B^2_{x_k} + B^2_{y_k} + B^2_{z_k}}\times dl_k}{\sum_k dl_k}
\end{equation}

\begin{figure*}[ht!]
\centering
\includegraphics[width=1\linewidth]{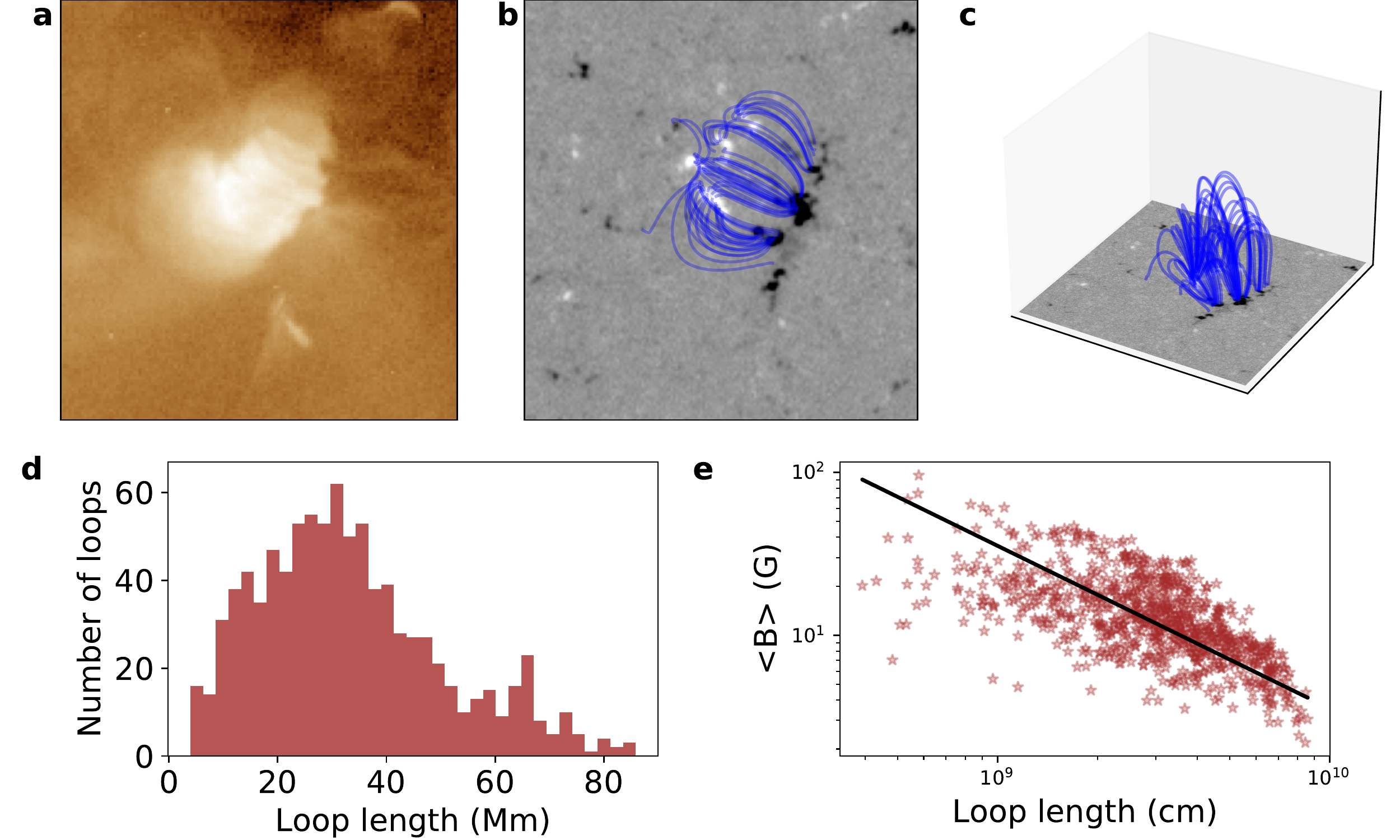}
\caption{Panel {\bf a} shows the representative image of an XBP as observed by AIA 193$\textup{\AA}$ channel. Panel {\bf b} shows the HMI magnetogram associated with the XBP and the blue curves are the plane-of-sky projected extrapolated field lines. A 3D view of the extrapolated field lines are shown in Panel {\bf c}. Panel {d} and {\bf e} shows the distribution of the loop lengths and magnetic field strength for all the loops associated with all the XBPs.}
\label{fig-extrapolation}
\end{figure*}

Figure~\ref{fig-extrapolation} shows the AIA 193$\textup\AA$ image of one of the XBP  in panel a and the corresponding HMI magnetogram in panel b. 
Extrapolated field lines projected onto the plane of the sky are overplotted in blue, and a view of the magnetic skeleton from a different angle is shown in panel~c. A qualitative comparison between the extrapolated field lines and the brightening visible in the AIA image reveals that the field extrapolation and line tracing adequately capture the geometry of the XBP.

We find the existence of 25 XBPs inside the chosen area, $F_v$. For each of these XBP, we extrapolate the magnetic field lines and estimate the loop lengths and magnetic field strengths. 
Figure~\ref{fig-extrapolation}d shows the distribution of all the loop lengths associated with all the XBPs and Figure~\ref{fig-extrapolation}e shows the distribution of their average magnetic field strength ($<B>$) along the loop length. 
The loop length distribution is found to peak near 30 Mm.
The average magnetic field is found to vary inversely with loop length.
The $<B> \propto L^{-1}$ relation is overplotted by a black solid line as a reference. 

\subsection{Heating function}\label{sec:heating_function}

Once the magnetic skeleton is created from the extrapolation, the loops need to be filled with heated plasma. We assume a spatially averaged volumetric heating function for each loop (erg cm$^{-3}$ s$^{-1}$) that has two parts: impulsive heating by transient events (nanoflares) and steady background heating. 
The background heating is chosen such that it can maintain a temperature of approximately 5.0$\times$10$^{5}$ K. The required heating rate can be estimated with a static equilibrium loop scaling law~\citep{Aschwanden_2005psci_book},
%, Rajhans_2021},  
\begin{equation}
H_{bkg}[erg cm^{-3} s^{-1}] \simeq \frac{2}{7} \Big(\frac{10}{9}\Big)^{\frac{7}{2}} k_0 \frac{\bar{T}^{\frac{7}{2}}}{L^2}
\end{equation}
Here, $k_0$=8.12$\times$10$^{-7}$ in cgs, $\bar{T}$ is the average temperature (in our case 5.0$\times$10$^{5}$ K) of the coronal part of the loop, which is related with the loop top temperature (T$_a$) as, $\bar{T} \approx 0.9T_a$~\citep{cargill_2012b}. 

Following~\cite{parker_1988} and ~\cite{klimchuk_2015RSPTA}, an impulsive event can occur with the release of stored magnetic energy that derives from slow photospheric driving. If $\theta$ is the angle between the stress component and potential component of the field, then the density of free magnetic energy available for heating is
 \begin{equation}\label{eq:max_heating_rate_avg}
    H = \frac{(tan(\theta) <B>)^2}{8\pi} (erg\hspace{0.1cm} cm^{-3}) 
\end{equation}
In the picture of tangled and twisted magnetic strands, $\theta$ is the tilt of the magnetic field from vertical at the base of the corona. It is sometimes referred to as the \textit{Parker angle}. It also corresponds to the misalignment half angle between adjacent strands at the time they start to reconnect. To satisfy the observed coronal heating energy requirements, $tan(\theta)=c$ should be in the range of $0.2-0.3$~\citep{parker_1988,klimchuk_2015RSPTA}. %($H_0^{min}$) 

Following \cite{klimchuck_2008ApJ}, \cite{cargill_2012b}, and \cite{Barnes_2016a}, we define the impulsive heating function in terms of a series of symmetric triangular heating profiles having a duration ($\tau$) of 100 s. The peak heating rate during an event ($H_0$) is randomly chosen between minimum ($H_0^{min}$) and  maximum ($H_0^{max}$) values that are loop dependent.
$H_0^{max}$ is determined from Equation~\ref{eq:max_heating_rate_avg}, so for $j^{th}$ loop of $i^{th}$ XBP it will be,
\begin{equation}\label{eq:max_heating_rate}
    H^{max}_{0_{ij}} = \frac{1}{\tau} \frac{(c <B>_{ij})^2}{8\pi} (erg\hspace{0.1cm} cm^{-3}\hspace{0.1cm}s^{-1}) 
\end{equation}
$H_0^{min}$ is taken to be  0.01 $\times$ $H_0^{max}$. 

As the free energy associated with a stressed loop is being released during an impulsive event, naturally, releasing a larger amount of energy causes a larger delay in storing enough energy that can be released during the next impulsive event.  Taking into account this important consequence, we assume that the delay time between the two consecutive events is proportional to the energy  of 1st event, i.e., the delay time between $(l-1)^{th}$ and $l^{th}$ event will be,
\begin{equation}
    d^l_{ij} = q \times H^{l-1}_{ij} 
\end{equation}
The value of the proportionality constant, $q$ is estimated by equating the average Poynting flux ($F$ in units of erg $cm^{-2}$ s$^{-1}$) associated with a loop with the average energy released by the impulsive events. This makes the above equation in the form:
\begin{equation}\label{eq-heating_delay} 
    d^l_{ij} = \frac{\tau L}{F} \times H^{l-1}_{ij}
\end{equation}

In the present study, we estimate $F$ by two different methods. The first method (called the $Constant-F$ model) assumes that all the loops associated with all the XBPs have the same average Poynting flux, which is calculated from the observed $DEM_\mathrm{XBP}$, as discussed in Section~\ref{sec:const_Fp}. The second method (called the $Variable-F$ model) assumes a different Poynting flux for each loop as discussed in Section~\ref{sec:var_Fp}. 

\subsubsection{Constant Poynting flux (Constant$-$F model)}\label{sec:const_Fp}
The total radiation loss rate ($\mathcal{R}$) from the solar atmosphere can be estimated using the observed line-of-sight EMD (in units of cm$^{-5}$) and radiation loss function ($\Lambda(T)$) as follows:
\begin{equation}\label{eq-rad_loss}
    \mathcal{R} = \sum_i  EMD(T_i) \hspace{0.1cm} \Lambda(T_i)
\end{equation}
Using the radiation loss function adopted in EBTEL~\citep{klimchuck_2008ApJ} and the observed EMD 
of the XBPs, we find that the average radiation loss for XBPs is 1.95$\times10^5$ erg cm$^{-2}$ s$^{-1}$.

The corona is cooled by both radiation and thermal conduction, the latter providing the energy that powers the radiation from the transition region. The heating Poynting flux must therefore balance the total radiative losses, including those from the transition region. The computed EMD in Equation~\ref{eq-rad_loss} does not extend below $logT = 5.6$ because the cooler values cannot be reliably measured. We must account for this missing radiation. In equilibrium loops, the radiative losses from the transition region are larger than those from the corona, and these losses are greatest in the lower and middle transition region. Following \cite{klimchuck_2008ApJ}, we take the total radiative losses in the loop to be 2-3 times larger than the coronal losses.
Thus, the average Poynting flux to each loop is,
\begin{equation}\label{eq-constant-F}
    F = g \times 1.95\times 10^5 \hspace{0.1cm} (erg \hspace{0.1cm} cm^{-2} \hspace{0.1cm} s^{-1})
\end{equation}
where $g$ is a constant in the range of 2 to 3.

\begin{figure*}[ht!]
\centering
\includegraphics[width=1\linewidth]{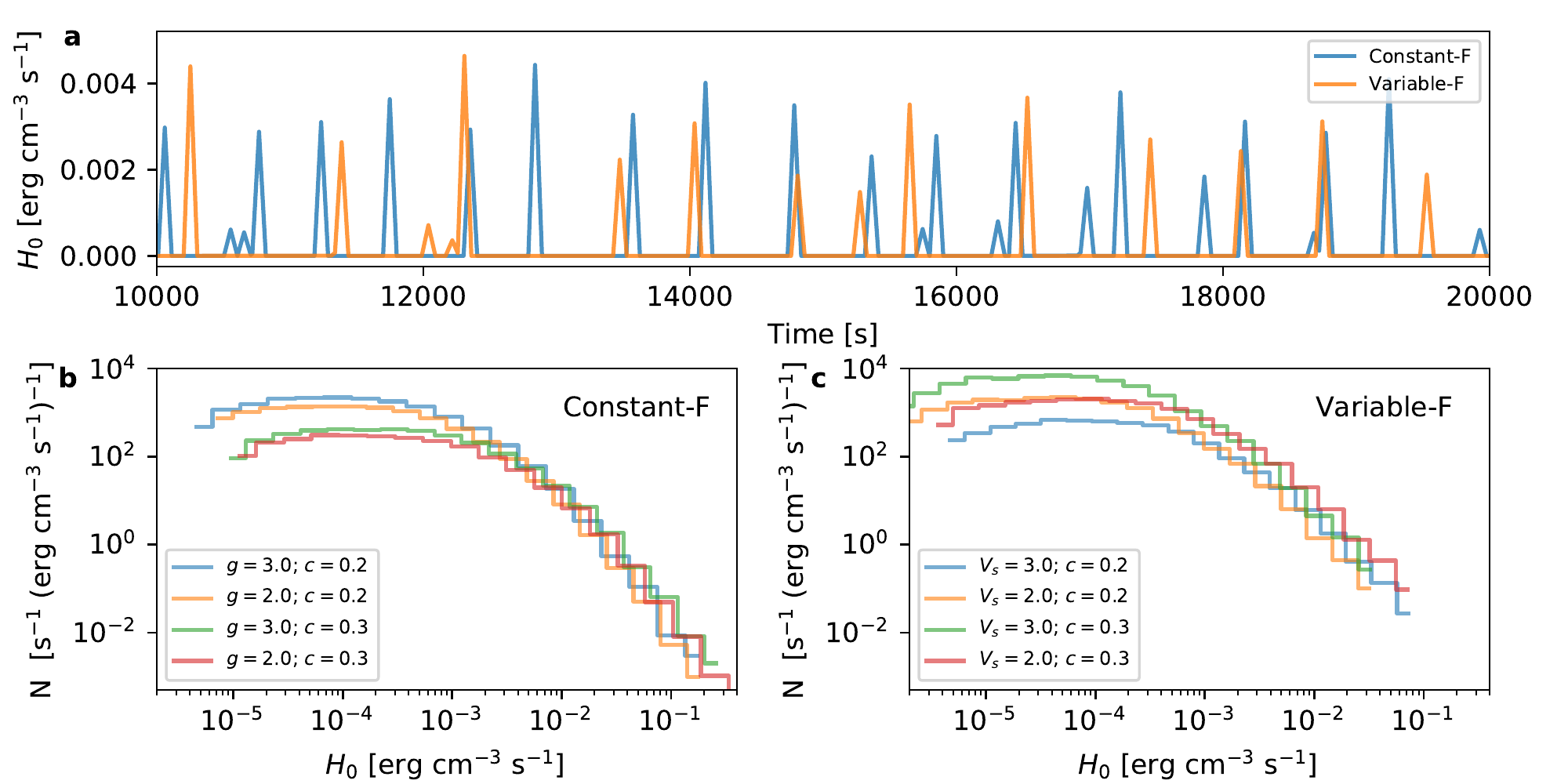}
\caption{Panel {\bf a} shows the representative heating function for a typical loop derived by using  Constant$-$F and Variable$-$F models. Panel {\bf b} and {\bf c} shows the heating frequency distribution of the events for Constant$-$F and Variable$-$F models respectively.}
\label{fig-heatingProfile}
\end{figure*}

Deriving the heating profile by combining Equations~\ref{eq:max_heating_rate_avg},~\ref{eq-heating_delay}, and \ref{eq-constant-F} (Constant$-$F model)
has four variable parameters; $L$, $<B>$, $c=tan(\theta)$, and $g$. 
Figure~\ref{fig-heatingProfile}a (blue line) shows the heating profile for  a loop of $L=$30 Mm, $<B>=$10 G, $g=$ 2.0, and $c=$ 0.25.
The $L$ and $B$ are derived from the magnetic modeling of the photospheric magnetogram (Section~\ref{sec-extrapolation}) while the exact values of $c$ and $g$ are unknown. However, we know their expected range for the coronal loops
as summarized in the first row of Table~\ref{table-I}. 
We have varied the values of $c$ and $g$ within their expected range to match the observation as discussed in Section~\ref{sec:sim_dem}.
Figure~\ref{fig-heatingProfile}b shows the distribution of the heating events associated with the loop distribution of XBPs (Figure~\ref{fig-extrapolation}d) for the combination of $c=$0.2, 0.3 and $g=$2.0, 3.0. 

\begin{deluxetable}{c c c c}
\tablecaption{Variable parameters and their expected range for the Constant$-$F and Variable$-$F models.}
\label{table-I}
\tablehead{
Model & $c$ = $\tan(\theta)$ & $g$ & $V_h$(Km/s)
% & & erg cm$^{-2}$ s$^{-1}$) & Km/s  
}
\startdata
$Constant-F$ & 0.2-0.3 & 2-3 & -- \\
$Variable-F$ & 0.2-0.3 & -- & 0.5-2.0
\enddata
\end{deluxetable}

\subsubsection{Variable Poynting flux (Variable$-$F model)}\label{sec:var_Fp}

It is to be noted that even if the photospheric driver flows are the same,  loops may not experience the same Poynting flux. This is because the field strengths vary from one loop to another. Following \cite{Klimchuck_2006SoPh} the expression for Poynting flux of individual loop can be written as

\begin{equation}\label{eq-F_base}
     F = -\frac{1}{4\pi}  V_h \tan(\theta)\hspace{0.1cm}(<B>)^2
 \end{equation}
In this particular case we have considered the loop to be non expanding, so that the field strength at the base of the corona is equal to the average field strength along the loop, $<B>$. $V_h$ is to be the horizontal speed of the flow that drives the field. However, this will not be the case if the loop expands with height. In such a situation, the cross sectional area over which the Poynting flux energy enters the loop is smaller than the area over which it heats the plasma. We can account for this difference with the modified expression (see Appendix~\ref{appecdix:avg_Pflux})

 \begin{equation}\label{eq-variavleF}
     F_{ij} = -\frac{1}{4\pi} V_h \tan(\theta)\hspace{0.1cm} B^{base}_{ij}<B>_{ij}
 \end{equation}
where $B^{base}$ is the magnetic field at the coronal base and the subscripts refer to the $j^{th}$ loop of $i^{th}$ XBP. 
We take the coronal base to be 2~Mm above the photosphere and we determine field strength there from the extrapolation. 

The heating profile of the Variable$-$F model is obtained by combining Equations~\ref{eq:max_heating_rate_avg},~\ref{eq-heating_delay}, and \ref{eq-variavleF}. It has five variable parameters: $L$, $<B>$, $B^{base}$,  $c=tan(\theta)$ and $V_h$. Figure~\ref{fig-heatingProfile}a (orange color) shows the derived heating profile  for a loop of $L=$30 Mm, $<B>=$10 G, $B^{base}=$15 G, $V_h=$ 1 Km/s, $c=$0.25.
Values of $L$, $<B>$ and $B^{base}$ are estimated from the magnetic modeling of the loops, whereas the exact values of $V_h$ and $c$ are unknown.
However, the expected range for these two variables is known  and summarized in the second row of Table~\ref{table-I}. For an example, 
Figure~\ref{fig-heatingProfile}c shows the distribution of the heating events corresponding to the loop distribution of XBPs (Figure~\ref{fig-extrapolation}d) for a combination of $V_s=$0.5, 1.5 and $c=$0.2, 0.3. 
This distribution is found to vary slightly according to the values of $V_h$, and $c$ and thus we have varied the values of $V_s$ and $c$ within their expected range to match the observation as discussed in Section~\ref{sec:sim_dem}.

\subsection{Simulated DEM}\label{sec:sim_dem}

Once the loop lengths and heating profiles for all the loops associated with all the XBPs are available, 
we run the EBTEL for individual loops in a parallel computing environment of a machine on 32 cores. 
Thus, in the simulation setup EBTEL is called multiple times associated with the different loops.
We simulate the evolution of the loops for the duration of 20000 s.
The estimated DEM of the transition region and coronal portion of the loops are stored for the last 7200 s of simulation time, similar to the observed DEM exposure time.
Combining the DEM of all the loops, we estimate the composite simulated DEM for all the XBPs.

\begin{figure}[ht!]
\centering
\includegraphics[width=1\linewidth]{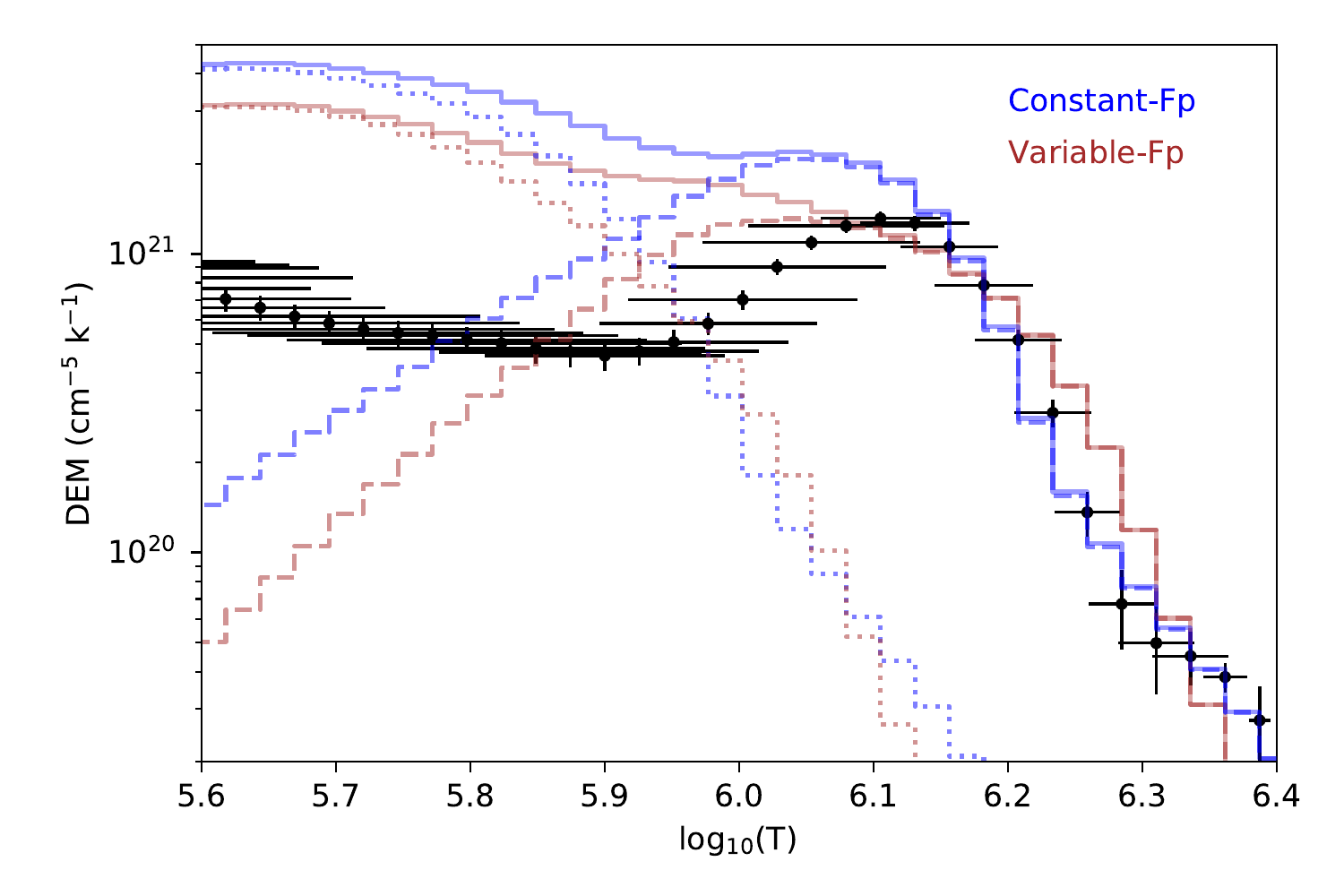}
\caption{Observed DEM (black errorbars) and simulated DEMs using Constant$-$F (blue color) and Variable$-$F model (brown color). The contribution of the transition region and coronal DEMs to the total simulated DEM (solid lines) are shown separately by dotted and dashed lines.}
\label{fig-SimulatedDEMs}
\end{figure}

\begin{figure}[ht!]
\centering
\includegraphics[width=1\linewidth]{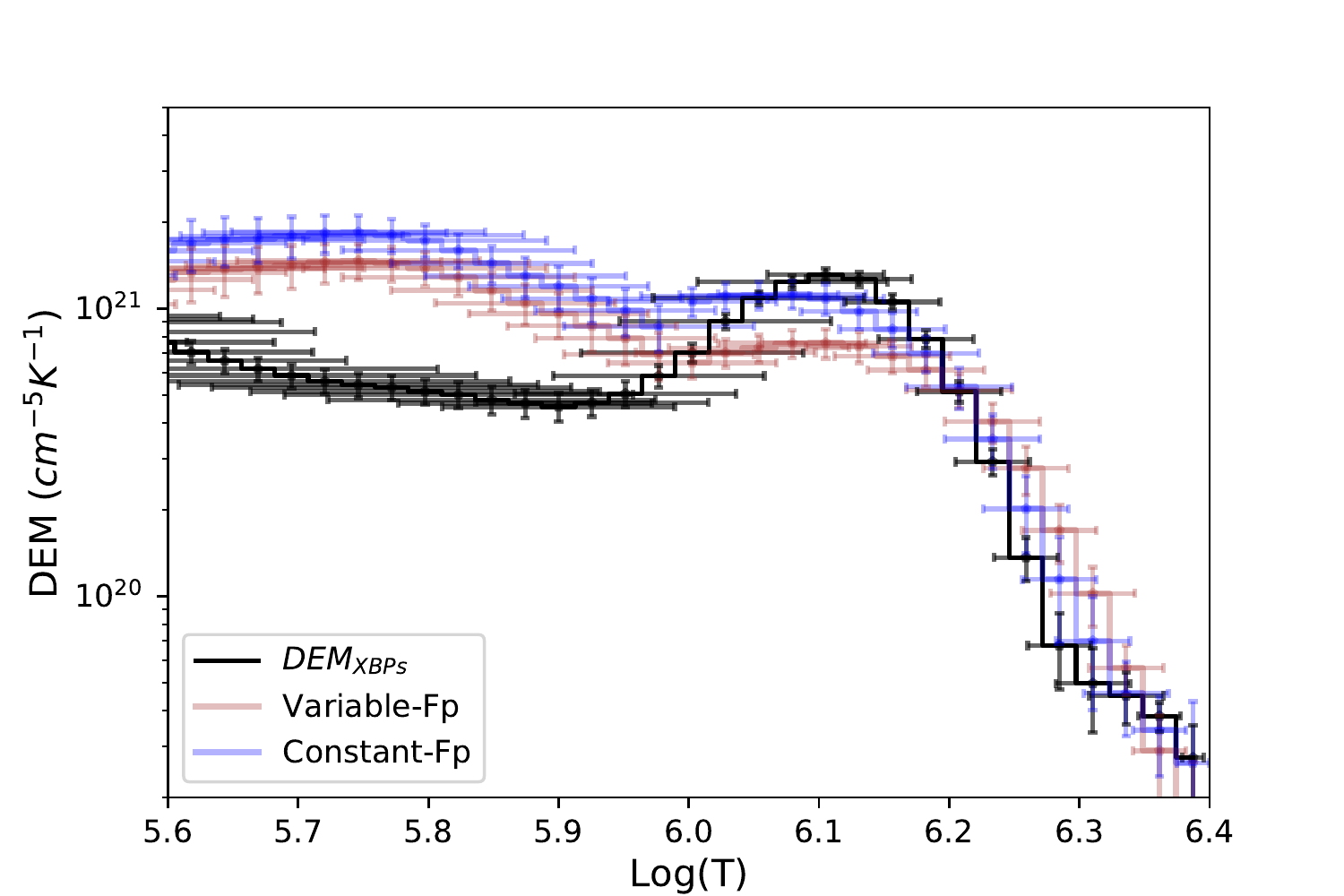}
\caption{Observed DEM of XBPs (black) compared with the recovered DEM obtain from the simulated AIA and XSM intensities from the simulated DEM shown in Figure~\ref{fig-SimulatedDEMs}}
\label{fig-SimulatedDEMs_HK_dem}
\end{figure}

The simulation setup is run multiple times by  varying the input parameters within their expected range (Table~\ref{table-I}) for both Constant$-$F (Section~\ref{sec:const_Fp}) and Variable$-$F (Section~\ref{sec:var_Fp}) models. We estimate the composite simulated DEM for each run and compare it with the observed DEM ($DEM_\mathrm{XBP}$).
The input parameters for which the simulated DEM well describes the observed DEM based on visual comparison are summarized in Table~\ref{table-II} and plotted in 
Figure~\ref{fig-SimulatedDEMs} (brown and blue colors).
The transition region and coronal portion of the simulated DEM are shown by dotted and dashed lines, respectively, whereas solid lines show the total DEMs.
Though both models predict emission at higher temperatures (logT $>$ 6.0) that is close to the observed emission,
they both predict emission at lower temperatures (logT $<$ 6.0)
that is $\sim$2 to 5 times too high. 
This low-temperature emission primarily comes from the transition region of the loops, which is poorly constrained by the AIA channels, as indicated by the larger error bars in the observed DEM.
Thus the recovered DEM at low temperature can underpredict the actual emission as demonstrated by~\cite{hannah_2008ApJ}.
To verify this scenario, we have predicted the AIA and XSM intensities from the simulated DEMs of the transition region and corona and recovered their DEMs using the \verb|HK_dem| method (Section~\ref{sec:XBP_dem}) by considering a typical 20$\%$ uncertainty in the simulated intensities.
We find that the recovered coronal emission  from the simulated intensities (logT$>$6.0 in Figure~\ref{fig-SimulatedDEMs_HK_dem}) 
matches well with the observed DEM. However, the recovered transition region DEM (logT$<$6.0) still shows a 2 to 3 times higher emission than the observed DEM.
Thus the deviation of the simulated and observed DEMs at a lower temperature is not only because of the observational uncertainty; rather, it indicates that the simulated transition region predicts a larger emission than the observed one -- details and possible explanations for this deviation are given in Section~\ref{sec-discussion}.

\begin{deluxetable}{c c c c}
\tablecaption{Best Suited parameters for the Constant$-$F and Variable$-$F models.}
\label{table-II}
\tablehead{
Model & $c$ = $\tan(\theta)$ & $g$ & $V_h$(Km/s)
% & & erg cm$^{-2}$ s$^{-1}$) & Km/s  
}
\startdata
$Constant-F$ & 0.21 & 2.47 & -- \\
$Variable-F$ & 0.21 & -- & 1.5
\enddata
\end{deluxetable}

%\subsection{Frequency distribution of impulsive events}
\subsection{Inferred frequency distribution}

\begin{figure*}[ht!]
\centering
\includegraphics[width=1\linewidth]{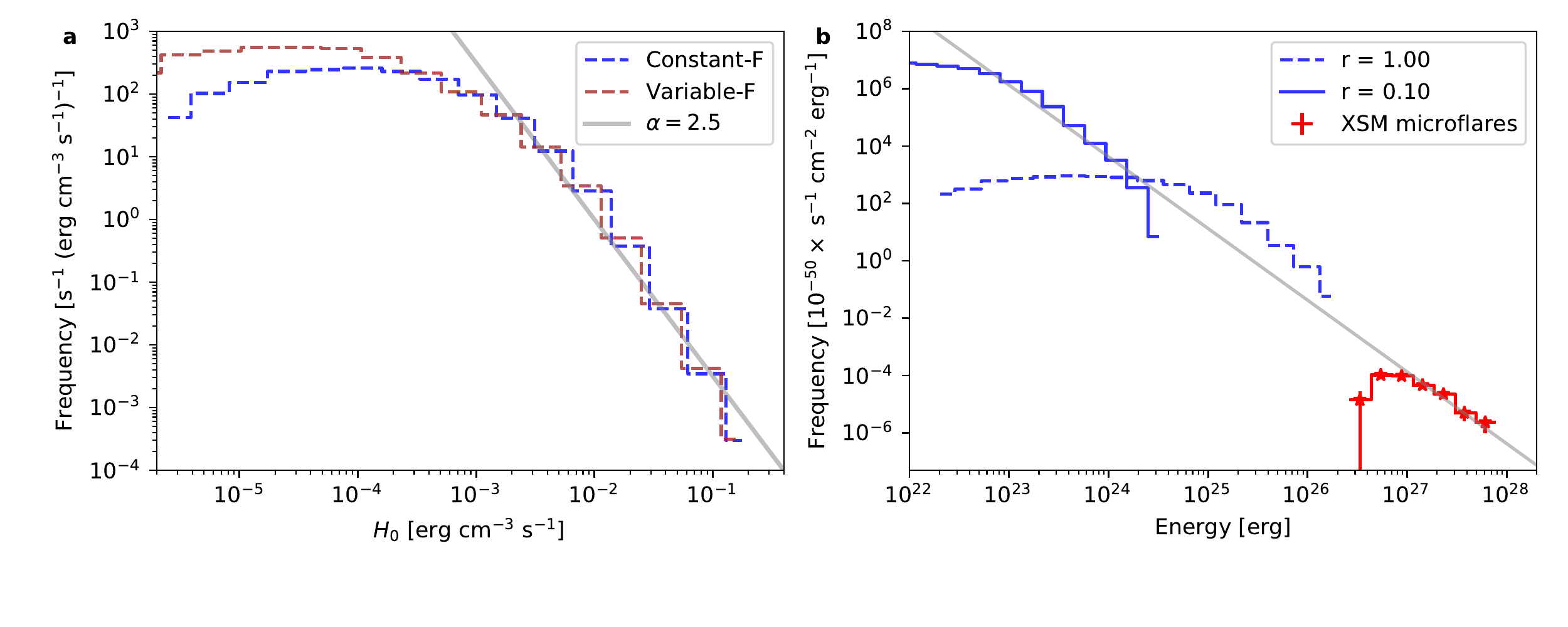}
\caption{Panel {\bf a}: Heating frequency distribution of Constant$-$F (blue) and Variable$-$F (brown) models. Grey line represents  a comparison power-law with a slope of $-2.5$. Panel {\bf b}: Energy distribution of the events for Constant$-$F model derived from the heating frequency shown in panel {\bf a}. The dashed and solid blue lines represent the energy distribution estimated by considering the a constant loop radius of 1 Mm and 0.1 Mm respectively. The grey solid line represents the power-law function of slope $-2.5$, which intersects with the XSM observed microflare frequency distribution at higher energies. 
}
\label{fig-FreqDist}
\end{figure*}

Figure~\ref{fig-FreqDist}a shows the frequency distribution of the impulsive event peak heating rates ($H_0$) for the model parameters that give the best match between simulated and observed DEMs (Table~\ref{table-II}).
At higher temperatures, the distribution is close to a power-law of slope -2.5, as indicated by the grey reference line.
We convert the heating rate distribution for the Constant$-$F model (blue dashed line in Figure~\ref{fig-FreqDist}a) to an energy distribution by integrating over the event duration and multiplying by the loop volume.
This is shown as a blue dashed line in Figure~\ref{fig-FreqDist}b, which is compared with the frequency distribution of the quiet Sun microflares as observed by XSM~\citep{xsm_microflares_2021}. During the minimum of solar cycle 24, these microflares are found to occur everywhere on the Sun outside the conventional AR and most of them are associated with the XBPs.
A comprehensive discussion on this is given in Section~\ref{sec-discussion}.

\section{ Discussion and Summary}\label{sec-discussion}
In the present work, we utilize the full disk observations of the Sun using AIA and XSM to derive the DEM of the disk-integrated Sun ($DEM_\mathrm{FullSun}$), X-ray emitting region ($DEM_\mathrm{XER}$), and X-ray bright points ($DEM_\mathrm{XBP}$) during the minimum of solar cycle 24. Our analysis suggests that in the absence of ARs, XBPs are the primary contributor to the total X-ray emission of the full Sun. Using hydrodynamic loop simulations we model the observed DEM of XBPs. The simulated DEM is then compared with the observed one. The primary findings of this paper are summarized below.\\

\paragraph{Quiet Sun coronal emission primarily consists of diffuse emission from the cold plasma and that of the XERs}
The disk integrated DEM ($DEM_\mathrm{FullSun}$; Figure~\ref{fig-EM_distribution}a) reveals a low temperature (around $1$ MK) peak along with an extended faint (approximately $2-3$ orders of magnitude less than the peak around $1$ MK) emission in the temperature range of $6.1<\log(T)<6.4$.  The peak is likely to be dominated by the emission from the cool quiet regions (also known as the diffuse corona) that occupies most of the solar disk.  
This peak emission is similar to earlier observations of the quiet Sun DEM (e.g., \citealp{Lanzafame_2005A&A,Brooks_2009,Giulio_2019A&A,Sylwester_2019SoPh}).
 On the other hand, the extended faint but high-temperature ($\log T > 6.1$) emission is expected to be dominated by the X-ray emitting regions (XER). 
Using the full disk images of AIA and corresponding observations from the XSM, we derive (Section~\ref{appendix-XbpDetec}) the $DEM$ of XERs ($DEM_\mathrm{XER}$; Figure~\ref{fig-EM_distribution}b). Successively EMD of the same is also derived. A comparison between the EMD of the full Sun (Figure~\ref{fig-EM_distribution}c) with that of the XERs (Figure~\ref{fig-EM_distribution}d) shows that the high-temperature components are similar, indicating that the high-temperature emission of the full Sun is primarily the source of the XERs.\\

\paragraph{During the quiet phase of the Sun, XBPs dominate the high-temperature emission observed by XSM} Figure~\ref{fig-XBP_detection} shows that in the absence of on-disk ARs, the emission from the limb and XBPs mostly constitute the emission of XERs. It is expected that 
%\red{(REF)} 
limb brightening is primarily due to the emission coming from a large volume of low-temperature plasma. To identify the emission of XBPs from that of the overall XERs we specifically derive (Section~\ref{sec-XBP_EMdist}) the DEM of XBPs that are present at the center of the solar disk (Figure~\ref{fig-HK_dem}a, blue color). $DEM_\mathrm{XER}$ \& $DEM_\mathrm{XBP}$ depict similar emission when $\log (T) > 6.1$, indicating that at this temperature range, XBPs primarily dominate the overall emission of XERs. Whereas, at low temperatures ($\log (T) < 6.1$), $DEM_\mathrm{XER}$  shows higher emission compared to $DEM_\mathrm{XBP}$, which may be due to the contribution from the limb brightening in $DEM_\mathrm{XER}$. 
For further verification, a typical $DEM$ of the limb is derived from the intensity of a small limb area selected from the full disk images taken by AIA and XRT (orange color in Figure~\ref{fig-HK_dem}). The limb DEM shows significant emission at low temperatures in the range, $\log(T) < 6.1$. \\

To quantify the emission coming from the quiet regions, XER, and XBPs, radiative fluxes from each of these regions are estimated. Equation~\ref{eq-rad_loss} is used for these estimations, while the inferred DEMs and the radiative loss function used in \cite{klimchuck_2008ApJ} are taken as input. 
Radiative fluxes are estimated in two temperature ranges; one is in the low-temperature emission ( $\mathcal{R}$($5.6 \leq \log T \leq 6.1$) ) while the other is in the high-temperature emission ( $\mathcal{R}$($6.1 \leq \log T \leq 6.4$) ) as is summarized in Table~\ref{table-III}.
The radiative flux of the full Sun, dominated by the quiet regions, is $\sim$0.9$\times$10$^5$ erg cm$^{-2}$ s$^{-1}$, which is close to the canonical quiet Sun value of ~\cite{Withbroe_1977}.
The high temperature component ($\log T > 6.1$) is almost an order of magnitude weaker than the cooler component.
XBPs account for the $63\%$ of the XER radiative flux at low temperatures ($\log T < 6.1$), while they contribute $85\%$ at high temperatures (log(T)$>$6.1).
This indicates that at high temperatures, most of the X-ray emissions observed by the XSM originate from the XBPs.\\

\begin{deluxetable}{c c c }
\tablecaption{Estimated radiative fluxes for full Sun, XER, and XBPs.}
\label{table-III}
\tablehead{
DEM used & $\mathcal{R}(5.6\leq logT\leq 6.1)$  & $\mathcal{R}(6.1\leq logT\leq 6.4)$\\
 & (erg cm$^{-2}$ s$^{-1}$) & (erg cm$^{-2}$ s$^{-1}$)
}
\startdata
$DEM_\mathrm{FullSun}$ & 0.78$\times$10$^5$ & 0.09$\times$10$^5$ \\
$DEM_\mathrm{XER}$ & 1.69$\times$10$^5$ & 1.01$\times$10$^5$\\
$DEM_\mathrm{XBP}$ & 1.08$\times$10$^5$ & 0.87$\times$10$^5$\\
\enddata
\end{deluxetable}

\paragraph{Results of the simulated XBPs agree well with the earlier findings }
Like active regions, XBPs consist of small-scale coronal loops~\citep{Madjarska_2019}. 
XBPs are found to be associated with bipolar regions (e.g., Figure~\ref{fig-extrapolation}b) on the photospheric magnetograms. Potential field extrapolation of these magnetograms (Section~\ref{sec-extrapolation}) provides the loop structures (e.g., Figure~\ref{fig-extrapolation}a,b,c) along with their length and magnetic field strength.
The composite distribution of loop lengths associated with all the XBPs (Figure~\ref{fig-extrapolation}d) shows a peak at around 30 Mm, which is much smaller than the typical loop lengths of the ARs (order of 10$^2$ Mm~\citealp{Aschwanden_2005psci_book}).
The average field strength (Figure~\ref{fig-extrapolation}e) of the loops is found to vary inversely with length to a power close to -1 (black solid line in Figure~\ref{fig-extrapolation}e), which is similar to the power law derived for AR loops~\citep{Mandrini_2000ApJ}. 

We used the EBTEL hydrodynamic model and observational constraints to simulate XBP loops (Section~\ref{sec-simulations}).
The loops were heated impulsively based on loop parameters derived from the extrapolations and observationally-based assumptions about the input Poynting flux (Section~\ref{fig-heatingProfile}).
Two assumptions were considered: an equal Poynting flux for all loops (Constant$-$F model; Section~\ref{sec:const_Fp}) and Poynting fluxes that are loop dependent (Variable$-$F model;  Section~\ref{sec:var_Fp}). 
Each of these models is associated with two unknown parameters ($c$, $g$ or $c$, $V_h$) as summarized in Table~\ref{table-I}.
Varying these parameters within their expected range obtained from earlier studies, we predicted the composite DEM of XBPs from the simulation and compared it with the observed one.
The input parameters that provide the best match are summarized in Table~\ref{table-III}. For the Variable$-$F model, the Parker angle ($\theta$) is found to be $\sim$12$^0$, which is close to the typical value of 10$^0$ for ARs~\citep{Klimchuck_2006SoPh}. The horizontal driver velocity is found to be $\sim$1.5 Km/s, which is close to the observable range.
%~\textcolor{red}{[REF]}. 
 For the Constant$-$F model, the total energy losses from the corona, including thermal conduction, are found to be 2.5 times the coronal radiative losses. This is also consistent with expectations based on 1D hydrodynamic simulations.\\
 
\paragraph{Simulated DEM agrees well with the observed one at high temperatures}
Figure~\ref{fig-SimulatedDEMs} shows a comparison between the observed and simulated DEMs.
The DEM obtained through the Constant$-$F model (blue line) agrees well with the observed DEM (black error bars) at high temperatures ($\log (T) > 6.1$), where most of the emission comes from the corona (dashed blue line).
The Variable$-$F model (brown line), on the other hand, slightly over predicts the DEMs at high temperatures. However, given the simplicity of the model, the differences of less than a factor of two are not significant. 

 The DEM is over predicted by factors of two to five at low temperatures.  This may be an artifact because of the instrument's broad temperature response functions. Such an artifact is expected to impact the observationally inferred DEM but not the simulated one. To check this, first we produce synthetic AIA and XSM intensities using the simulated DEM. These synthetic intensities are then further utilized to reconstruct a new DEM, as is done in case of actual observations (Figure~\ref{fig-SimulatedDEMs_HK_dem}). Though the discrepancy is reduced to a factor of two to three, the newly obtained DEM still shows high emission at low temperatures.
 Predicting excess emission at transition region temperatures ($ \log T < 6.0$) is fairly common (e.g., \citealp{Warren_2008ApJ}) in loop simulations. It must be mentioned that frequent chromospheric jets (such as spicules) are responsible for absorbing a significant amount (by a factor of $2-3$) of transition region emission~\citep{Pontieu_2009ApJ}, causing a lower emission in the corresponding temperatures. Another possibility is that the emitting area of the transition region is reduced because loops are substantially constricted at their base due to the clumpiness of the magnetic field in the photosphere and the rapid transition from high-$\beta$ to low-$\beta$ conditions~\citep{Warren_2010ApJ...711..228W,Cargill_2022MNRAS}.

\paragraph{Simulations suggest a stiff power-law slope for the smaller flares}
When the heating rate is high, i.e., $H_0 > 10^{-3}$ erg cm$^{-3}$ s$^{-1}$, the composite frequency distribution of all the simulated loops maintains a power-law slope close to $-2.5$ (Figure~\ref{fig-FreqDist}a). Such a slope indicates that the combined energy of small scale impulsive events or nanoflares have more energy compared to their bigger counterparts, namely, flares~\citep{parker_1988} and microflares. Earlier observations also suggest~\citep{xsm_microflares_2021} that in the quiet Sun the bigger events occur only occasionally, with an average frequency of $\sim$1.8/days. The composite frequency distribution becomes flatter towards the lower heating rate ($H_0$). 

Integrating the heating distribution (Figure~\ref{fig-FreqDist}a) over the duration of the event and multiplying by the loop volume, we obtain the typical flare energy distribution, shown by the dashed blue line in Figure~\ref{fig-FreqDist}b. Extrapolating the distribution to higher energies disagrees with the earlier quiet Sun microflare distribution (red points) observed by XSM~\citep{xsm_microflares_2021}. We also note that there is an excess of nanoflares at the lower energies. Our XBP models assume that loops have a radius of $1$ Mm. Had we assumed a smaller radius of, say, $0.1$ Mm, the energy per nanoflare would be reduced by a factor of $100$ and there would be 100 times more loops in each XBP. The net effect is to shift the energy distribution to the left and upward, as shown by the solid blue curve in the figure. Now the nanoflares and microflares are both nicely explained by a single power-law distribution, suggesting a similar physical origin. We note that a coronal radius of $0.1$ Mm is consistent with the expected size of elemental magnetic strands~\citep{klimchuk_2015RSPTA}.\\ 

Carrying out a prolonged investigation of the quiet solar corona by separating out the contributions from its various emission components often becomes challenging in the Sun-as-a-star mode observations. Such an opportunity was provided by excellent observations of the quiet solar corona in the absence of any AR during the minimum of solar cycle 24.
Estimating the DEM and, subsequently, the radiation flux of the quiet corona, X-ray emitting regions, and XBPs, we found most of the quiet or diffuse corona emit at low temperatures ($ \log T < 6.1$). In contrast, most emission above $ \log T=6.1$ originate from XBPs. 
DEM of the modeled XBPs indicates that XBP heating is likely to be maintained by the small-scale nanoflares. They originate through the release of stored magnetic energy within the stressed magnetic loops.  
Along with the sophisticated modelling efforts, spatially resolved spectroscopic observations with an instrument capable of both low and high-temperature diagnostics are essential for comprehending the heating of small scale loops. An imaging spectroscopic instrument with good spatial and energy resolution in the X-ray energy range (e.g., below 1 KeV to 15 keV) could be beneficial in this context.

\acknowledgments{
We acknowledge the use of data from the Solar X-ray Monitor (XSM) on board the Chandrayaan-2 mission of the Indian Space Research Organisation (ISRO), archived at the Indian Space Science Data Centre (ISSDC). 
The XSM was developed by Physical Research Laboratory (PRL) with support from various ISRO centers.
We thank various facilities and the technical teams from all contributing institutes 
and Chandrayaan-2 project, mission operations, and ground segment teams for their support.
Research at PRL is supported by the Department of Space, Govt. of India. J.A.K. was supported by the Internal Scientist Funding Model (competed work package program) at Goddard Space Flight Center.
We acknowledge the support from Royal Society through the international exchanges grant No. IES{\textbackslash}R2{\textbackslash}170199.
GDZ and HEM acknowledge support from STFC (UK) via the consolidated grant to the atomic astrophysics group
at DAMTP, University of Cambridge (ST{\textbackslash}T000481{\textbackslash}1).
}

\renewcommand\thefigure{\thesection.\arabic{figure}}
\setcounter{figure}{0}

\appendix 
\section{XSM and AIA Temperature Response}\label{XSM_Tresp}

The XSM temperature responses are constructed from individual isothermal emission models over a logarithmic grid ($\delta$(LogT) = 0.03) of temperatures (T) from 0.5 MK to 50 MK. We use the XSPEC~\citep{ref-xspec} local model, \verb|chisoth|~\citep{biswajit_2021} for the estimation of the isothermal emission spectrum at each temperature grid. 
As we are interested in the analysis of the quiet solar corona, we adopt the quiet sun elemental abundances from ~\cite{xsm_XBP_abundance_2021}.  
At the time of model calculation,
we  use the energy response (RMF) function of the XSM. However, as the XSM effective area is varying with time, the time-varying effective area file (ARF) is used for the observation duration. These RMF and ARF are folded with the synthetic photon spectrum and produce the synthetic count spectrum of XSM in the units of Counts s$^{-1}$ keV$^{-1}$ for an emission measure of 10$^{46}$ cm$^{-3}$. 
We multiply the output spectrum by a factor  \textit{(10$^{-46}$ $\times$ energybin)},  further multiply by the emitting plasma \textit{area} (e.g., the total area of X-ray emitting regions) 
to convert it into the units of Counts cm$^5$  s$^{-1}$ for a unit emission measure. 
To get the temperature response from these synthetic spectra at different temperature grids, we integrate the average counts over the dynamic energy bins  of 1.29-1.45 keV, 1.45-1.75 keV, 1.72-1.95 keV, and 1.95-2.5 keV.
Thus we have a matrix of plasma temperatures and the XSM re-bind energy band for which we have the predicted count rates per unit emission measure. 

\begin{figure}[ht!]
\centering
\includegraphics[width=1\linewidth]{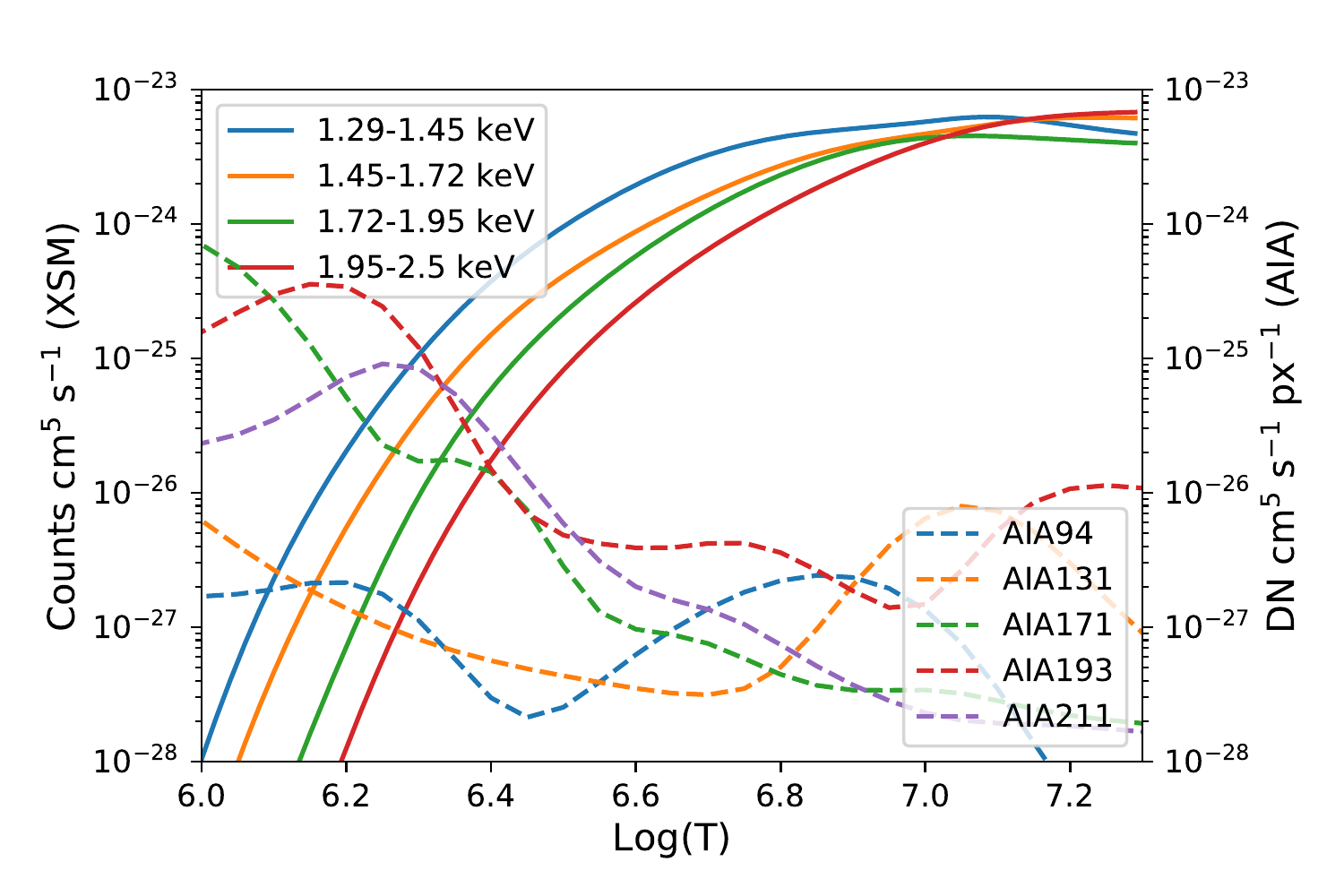}
\caption{Temperature response functions for XSM (solid lines in unit Counts cm$^5$ s$^{-1}$) and AIA (dashed lines in unit DN cm$^5$ s$^-1$ px$^-1$)
}
\label{fig-AIA_XSM_Tresp}
\end{figure}

The temperature response functions for the SDO/AIA EUV channels are obtained using the standard  routine 
\verb|aia_get_response.pro| available within the SSW. We use the same quiet Sun abundances obtained from ~\cite{xsm_XBP_abundance_2021} with CHIANTI version 10 and
adopted the latest calibrations, which incorporate the time-dependent corrections in the effective area.
Figure~\ref{fig-AIA_XSM_Tresp} shows the temperature response functions for the AIA (dashed lines) channels along with the four XSM channels (solid lines) for integrated emission of QS-1.
It should be noticed that the XSM temperature sensitivity starts to increase above 2 MK, whereas the AIA sensitivity
starts dropping at those temperatures. Furthermore, XSM also shows a good overlap in the temperature sensitivity with the AIA. 
Thus, the combined DEM derived by XSM and AIA constrains both low and high-temperature emissions.

\section{Average Poynting flux}\label{appecdix:avg_Pflux}

It was argued earlier (Section~\ref{sec:var_Fp}) that in case of non expanding loops one can evaluate the Poynting flux with Equation~\ref{eq-F_base}. However, this gets modified when we consider expanding loops. The Poynting flux at the base of the corona can be written as 

\begin{equation}\label{eq-F_phot}
     F^{base} = -\frac{1}{4\pi}  V_h \tan(\theta)\hspace{0.1cm}(B^{base})^2
 \end{equation}
 
 Here, $B^{base}$ is the magnetic field strength at the base of the corona. Let us further consider $A^{base}$ be the area of the loop at the same location. If $<A>$ and $<B>$ are the average area and magnetic field strengths along the loop, respectively, then following the conservation of magnetic flux one may write
\begin{equation}\label{eq-const_flux}
    B^{base}A^{base} = <B><A>
\end{equation}
Also, from the conservation of energy,
\begin{equation}\label{eq-constE}
    F^{base}A^{base} = Q L <A>
\end{equation}
where Q is the volumetric heating rate and L is the halflength of the loop. Combining Equation~\ref{eq-F_phot}, \ref{eq-const_flux} \& \ref{eq-constE};
\begin{equation}
    Q = -\frac{1}{4\pi}  V_h \tan(\theta)\hspace{0.1cm}\frac{B^{base}<B>}{L}
\end{equation}
This brings out the expression for the Pointing flux of expanding loop to be,
\begin{equation}
    F = -\frac{1}{4\pi}  V_h \tan(\theta)\hspace{0.1cm}B^{base}<B>
\end{equation}

\newpage
\bibliography{myref}   % bibliography data in report.bib
\bibliographystyle{apalike} % makes bibtex use apalike.bst

\end{document}